\documentclass[aps,prb,groupedaddress,showpacs,twocolumn,superscriptaddress]{revtex4}
\usepackage{graphicx}
\usepackage{amsmath}
\usepackage{amssymb}
\usepackage{bbm}
\usepackage{xspace}
\usepackage{color}
\usepackage{footnote}
\usepackage{hyperref}
\usepackage{subfigure}
\usepackage{scalerel}
\usepackage{stackengine,wasysym}
\usepackage{dsfont}
\usepackage{scrextend}
\usepackage{multirow}
\usepackage{hhline}
\definecolor{green1}{RGB}{00,100,00}
\definecolor{orange}{rgb}{1.0, 0.49, 0.0}

\newcommand{\fig}[1]{Fig.\thinspace{}\ref{#1}}
\newcommand{\figs}[1]{Figs.\thinspace{}\ref{#1}}
\newcommand{\eq}[1]{Eq.\thinspace{}\eqref{#1}}
\newcommand{\eqs}[1]{Eqs.\thinspace{}\eqref{#1}}
\newcommand{\se}{Sec.\@\xspace}
\newcommand{\app}{App.\@\xspace}
\newcommand{\tcite}[1]{Ref.~\onlinecite{#1}}
\newcommand{\tcites}[1]{Refs.~\onlinecite{#1}}

\newcommand{\teps}{\tilde \varepsilon}

\newcommand{\rre}{\Re{}\,}

\newcommand{\nag}{{\phantom{\dagger}}}

\def\ket#1{\mathinner{|{#1}\rangle}}
\def\bra#1{\mathinner{\langle{#1}|}}
\def\braket#1{\mathinner{\langle{#1}\rangle}}
\def\Braket#1{\mathinner{\left<{#1}\right>}}
\newcommand\varpm{\mathbin{\vcenter{\hbox{%
  \oalign{\hfil$\scriptstyle+$\hfil\cr
          \noalign{\kern-.3ex}
          $\scriptscriptstyle({-})$\cr}%
}}}}
\newcommand\varmp{\mathbin{\vcenter{\hbox{%
  \oalign{$\scriptstyle({+})$\cr
          \noalign{\kern-.3ex}
          \hfil$\scriptscriptstyle-$\hfil\cr}%
}}}}

\newcommand{\sdots}{\scalebox{0.5}{\textbf{\ldots}\,}}

\DeclareMathOperator{\Tr}{Tr}

\newcommand{\almostfulldef}{``full''\xspace}
\newcommand{\almostfull}{full\xspace}
\newcommand{\almostemptydef}{``empty''\xspace}
\newcommand{\almostempty}{empty\xspace}
\newcommand{\nontildedef}{``nontilde''\xspace}
\newcommand{\nontilde}{nontilde\xspace}

\newcommand{\pw}{0.85}
\begin{document}

\title{Pseudogap Anderson impurity model out of equilibrium:\\ A master equation tensor network approach}

\author{Delia M. Fugger}
\affiliation{Institute of Theoretical and Computational Physics, Graz University of Technology, Petersgasse 16/II, 8010 Graz, Austria}

\author{Daniel Bauernfeind}
\affiliation{Institute of Theoretical and Computational Physics, Graz University of Technology, Petersgasse 16/II, 8010 Graz, Austria}
\affiliation{Center for Computational Quantum Physics, Flatiron Institute, 162 Fifth Avenue, New York 10010, New York}

\author{Max E. Sorantin}
\affiliation{Institute of Theoretical and Computational Physics, Graz University of Technology, Petersgasse 16/II, 8010 Graz, Austria}

\author{Enrico Arrigoni}
\email{arrigoni@tugraz.at}
\affiliation{Institute of Theoretical and Computational Physics, Graz University of Technology, Petersgasse 16/II, 8010 Graz, Austria}

\begin{abstract}

We study equilibrium and nonequilibrium properties of the single-impurity Anderson model with a power-law pseudogap in the density of states.
In equilibrium, the model is known to display  a quantum phase transition from a generalized Kondo to a local moment phase.
In the present work, we  focus on the extension of these phases beyond equilibrium, i.e. under the influence of a bias voltage.
Within the auxiliary master equation approach combined with a scheme based on matrix product states (MPS) we are able to directly address the current-carrying steady state.
Starting with the equilibrium situation, we first corroborate our results by comparing with a direct numerical evaluation of ground state spectral properties of the system by MPS.
Here, a scheme to locate the phase boundary by extrapolating the power-law exponent of the self energy produces a very good agreement with previous results obtained by the numerical renormalization group.
Our nonequilibrium study as a function of the applied bias voltage is then carried out for two  points on either side of the phase boundary. 
In the Kondo regime the resonance in the spectral function is splitted as a function of the increasing bias voltage. 
The local moment regime, instead, displays a dip in the spectrum near the position of the chemical potentials. 
Similar features are observed in the corresponding self energies. 
The Kondo split peaks approximately obey a power-law behavior as a function of frequency, whose exponents depend only slightly on voltage. 
Finally, the differential conductance in the Kondo regime shows a peculiar maximum at finite voltages, whose height, however, is below the accuracy level. 

\end{abstract}

\maketitle
\section{Introduction}
\label{sec:intro}

The single-impurity Anderson model (SIAM) was originally introduced to address the properties of metals with dilute magnetic impurities, which displayed an unusual resistance minimum upon decreasing the temperature.\cite{ha.bo.34,sa.co.64}
This effect was termed Kondo effect and it was traced down to the formation of a highly entangled ground state of the model, namely, a singlet state between the localized impurity electron and the conduction electrons of the host metal screening the impurity spin. 
This has important consequences, such as the existence of a regime, in which physical quantities obey a set of universal scaling laws, which are independent of the microscopic details of the actual physical system.
In the Kondo regime, i.e. well below the so-called Kondo temperature $T_K$, the SIAM also behaves as a Fermi liquid.
Above this energy scale, the impurity spin is no longer screened and the model displays a crossover from the Kondo to a local moment (LM) regime.
In the impurity spectrum, this crossover is signaled by a strong suppression and broadening of the Kondo resonance, which, however, never completely vanishes. 
It is important to mention that there is no true quantum phase transition (QPT) in this model.\cite{wils,hews}

In the last decades, the SIAM has drawn renewed interest, due to its application in dynamical mean-field theory (DMFT), which has paved the way to understand the properties of a variety of correlated materials.\cite{ge.ko.96,voll.10}
It has further drawn attention, due to its capability to capture the physics of quantum dots, which can now be faithfully fabricated in the laboratory.\cite{go.go.98,fu.do.18}
These applications have in common that they usually deal with a structured density of states (DOS) of the host material, instead of a flat one, as in the original model. 
In contrast to metals, materials with a band gap %
cannot (fully) display the Kondo effect, since a finite DOS in a small region around the Fermi energy is crucial for its occurrence.
However, there are also materials, such as peculiar semiconductors and superconductors,\cite{vo.pa.85,si.ue.91} that display a pseudogap (PSG), i.e.~a DOS vanishing exactly at the Fermi energy with a certain power-law $\propto |\omega|^r$, but remaining finite, elsewhere. 
For this type of materials, the interaction of band fermions with a magnetic impurity produces more intriguing effects.\cite{wi.fr.90}
The corresponding PSG SIAM displays a rich zero-temperature phase diagram. In particular, for $0 < r < \tfrac{1}{2}$ it features a second-order QPT\,\cite{vojt.06} from a Kondo screened phase to a LM phase depending on the interplay between the power-law exponent $r$, the interaction and hybridization strengths. %
In this model, the depletion of host states at the Fermi energy prevents the impurity spin from being entirely screened by the conduction electrons.
As a consequence, the PSG SIAM does not behave as an ordinary Fermi liquid in the Kondo phase. Its behavior is captured by a natural, but non-trivial generalization of Fermi liquid theory, and the phase is referred to as a generalized Kondo (GK) phase.
Also in this case, a Kondo scale and a set of universal laws for the physical observables in terms of this scale is found, which is  distinct from the ordinary SIAM.\cite{wi.fr.90,bo.hi.92,ca.fr.96,ca.fr.97,vo.bu.01,ch.ja.95,go.in.96,inge.96,bu.pr.97,go.in.98,lo.gl.00,gl.lo.00,bu.gl.00,in.si.02,gl.lo.03,vo.fr.04,fr.vo.04,fr.fl.06,le.bu.05,bu.co.08,gl.ki.11,ka.ma.12,fr.vo.13,aono.13}

In this paper, we are interested in understanding the properties of the PSG SIAM, when a bias voltage $\phi$ is applied to drive the system out of equilibrium.\cite{mi.ta.06,di.mi.08,ho.gr.08,ch.hu.09,ki.si.09_1,ta.wi.10,ri.si.13,si.hu.13}
This model has been studied in previous works as well with different degrees of approximation and addressing different physical questions.
In \tcite{schi.12}, the PSG SIAM was studied after a local quench within a time-dependent Gutzwiller variational scheme.
The author found that the system thermalizes within the GK phase, but when quenching across the phase boundary, thermalization does not occur, and a highly nontrivial dynamical behavior is observed. 
\tcites{ch.zh.12,ri.za.15} both deal with universal scaling in the nonequilibrium steady state of the PSG Kondo model, employing variants of the renormalization group and large-$N$ techniques, respectively.
In the LM phase, close to the phase boundary, \tcite{ch.zh.12} reports universal scaling of the differential conductance, spin susceptibility and conduction electron $T$ matrix as a function of $\phi/T_K$. %
In \tcite{ri.za.15}, on the other hand, it was discovered that the differential conductance, spin susceptibility and Kondo-singlet strength, 
reproduce their equilibrium behavior in the scaling regimes of the fixed points of the model, when expressed in terms of a fixed-point specific effective temperature $T_\mathrm{eff}$. %
\tcite{ha.ko.15}, in contrast, focuses on the steady state impurity spectrum and differential conductance, the main quantities that also we are interested in within this work. 
Employing second-order perturbation theory, the authors find a cusp or dip structure in the impurity spectrum in the GK and LM phase, respectively, when a finite bias voltage is applied. 
However, in \tcite{ha.ko.15}, when increasing the bias voltage, these structures remain located at zero frequency and no splitting occurs.  
According to the authors, this is, because the system is not in the limit of large interaction strength. 
The results of our present work, while confirming the presence of these features, present a different scenario: the structures do split as a function of voltage.
One should point out that, while our calculations are carried out for values of the parameters very close to the ones used in \tcite{ha.ko.15}, there is a difference in the way the DOS pseudogap evolves as a function of voltage.
More specifically, in \tcite{ha.ko.15} the pseudogap is fixed at zero frequency also at finite bias voltages and only the chemical potentials are shifted by $\pm \phi/2$. 
In our work, on the other hand, we pin the pseudogap of each lead to the position of the respective chemical potential. 

We study the PSG SIAM out of equilibrium by an approach which is non-perturbative, neither in the interaction nor in the hybridization. 
Specifically, we employ the auxiliary master equation approach (AMEA),\cite{ar.kn.13,do.nu.14,do.so.17,fu.do.18} in which the nonequilibrium bath is accurately represented by an open quantum system, whose many-body dynamics is controlled by a Lindblad equation. 
The latter is solved by an efficient matrix product states (MPS) formulation.
We start by a benchmark of the approach in equilibrium. 
Here, in particular, we exploit the power-law exponent of the self energy to find the boundary between the GK and the LM phase.
We then carry on with a qualitative analysis of the structure of the spectral function and the self energy out of equilibrium in both the GK and LM regimes.
Besides these qualitative aspects, we try to fit a power-law behavior to these quantities in a region around the chemical potentials and investigate, how the corresponding power-law exponents evolve upon increasing the bias voltage. 
Finally, we address the behavior of the differential conductance in dependence of the bias voltage.
Our method is numerically exact, the main limitation being the fact that the pseudogap exponent in the bath DOS can be reproduced  only with a limited resolution. 
Therefore, we are also limited in the maximum bias voltage, in which our power-law analysis makes sense. 

This work is organized as follows:
In \se \ref{sec:modelmethod} the model and the solution method are described, starting with the model in \se \ref{subsec:model}, followed by a small overview about nonequilibrium Green's functions in \se \ref{subsec:neq_gfs} and a description of the auxiliary master equation approach in \se \ref{sec:amea}. 
Specifically, we present the Lindblad equation in \se \ref{subsubsec:Lindblad}, discuss the mapping to the auxiliary system in \se \ref{subsubsec:Mapping} and introduce the novel MPS scheme in \se \ref{subsubsec:MPS}.
\se \ref{subsubsec:phys_vs_aux} presents remarks about physical and auxiliary quantities.
\se \ref{sec:results} contains the results of this work, in particular, the results of the fit, \se \ref{subsec:results_mapping}, and the ones of the many-body solution in equilibrium, \se \ref{subsubsec:equilibrium}, as well as out of equilibrium, \se \ref{subsubsec:nonequilibrium}.
A discussion of the results obtained is found in \se \ref{sec:conclusion}.
\section{Model and Method}
\label{sec:modelmethod}
\subsection{Model}
\label{subsec:model}
We study the single-impurity Anderson model (SIAM) in as well as out of equilibrium with electronic leads displaying a power-law pseudogap (PSG) in the density of states (DOS).
Throughout this paper we use units of $\hbar=e=k_B=1$.
The model is described by the following Hamiltonian,
\begin{equation}
\label{h}
 H =  H_{\mathrm{imp}} +  H_{\mathrm{leads}} + H_{\mathrm{coup}} \,.
\end{equation}
$H_{\mathrm{imp}}$ is the Hamiltonian of the impurity. 
It is a single-site Hubbard Hamiltonian with on-site interaction $U$, accounting for the Coulomb repulsion between electrons, and on-site energy $\varepsilon_{f} = -\tfrac{U}{2}$, producing particle-hole (PH) symmetry,
\begin{equation}
 H_{\mathrm{imp}} =  \sum_{\sigma} \varepsilon_{f} f^\dagger_{\sigma} f_{\sigma}^\nag + U n_{f\uparrow}n_{f\downarrow} \,.
\end{equation}
$f^{\dagger}_{\sigma} / f^\nag_{\sigma}$ creates/annihilates an impurity electron with spin $\sigma \in \{\uparrow,\downarrow\}$ and $n_{f\sigma} = f^{\dagger}_{\sigma}f^\nag_{\sigma}$ is the corresponding particle-number operator.
$H_{\mathrm{leads}}$ is the Hamiltonian of the left and right lead, $\lambda \in \{L,R\}$,  
\begin{equation}
 H_{\mathrm{leads}} = \sum_{\lambda k \sigma} \varepsilon_{\lambda k}^\nag d^\dagger_{\lambda k \sigma}d^\nag_{\lambda k \sigma}\,. 
\end{equation}
It describes a continuum ($N\to\infty$) of noninteracting energy levels $\varepsilon_{\lambda k} = \varepsilon_{k} + \teps_\lambda $ rigidly shifted symmetrically by half the bias voltage $\phi$, so that $\teps_\lambda = \pm \frac{\phi}{2}$.
$d^\dagger_{\lambda k \sigma}/d^\nag_{\lambda k \sigma}$ are the corresponding creation/annihilation operators. 
Finally, 
\begin{equation}
 H_{\mathrm{coup}} = \frac{t'}{\sqrt{N}} \sum_{\lambda k \sigma} \left( d^\dagger_{\lambda k \sigma}f^\nag_{\sigma} + f^\dagger_{\sigma}d^\nag_{\lambda k \sigma} \right) %
\end{equation}
is the Hamiltonian that describes the coupling of the impurity to the leads via hoppings $t^\prime$.

We assume that the leads are initially decoupled ($t'=0$) and in equilibrium at the same temperature $T$ and chemical potentials $\mu_\lambda$ with an occupation given by the Fermi function,   
\begin{equation}
 \label{eq:ffermi}
 f_\lambda(\varepsilon,T) = \frac{1}{1+\exp\left( \frac{\varepsilon - \mu_\lambda}{T} \right)}\,.          
\end{equation}
Requiring the (asymptotical) particle density of each lead to be independent of $\phi$ amounts to setting $\mu_\lambda = \teps_\lambda$.

The leads have a power-law PSG DOS at $\mu_\lambda$, which we describe with the retarded hybridization functions, 
\begin{align}
 \Im \Delta^R_\lambda(\omega) &= - \pi \frac{{t^\prime}^2}{N} \sum_k \delta(\omega-\epsilon_{\lambda k}) \nonumber \\
 &= -\frac{\Gamma}{2} e^{-\gamma (\omega-\teps_\lambda)^2} |\omega-\teps_\lambda|^r \label{eq:im_hyb_R} \,,
\end{align}
whose symmetric forms produce a PH symmetric occupation of the leads.
Here, $\Gamma$ is the hybridization strength and $\gamma>0$ is used to fix the bandwidth.\footnote{
A Heaviside step function would also fix the bandwidth without distorting the power-law. 
We choose the exponential, because AMEA performs better for smooth hybridization functions.
}
The Keldysh hybridization functions are fixed by the fluctuation-dissipation theorem, 
\begin{equation}
 \label{eq:hyb_K}
 \Delta^K_\lambda(\omega) = 2i\, \left(1 - 2 f_\lambda(\omega,T)\right) \Im \Delta^R_\lambda(\omega) \,,
\end{equation}
and the total hybridization function at the impurity, accounting for both the left and the right lead, $\Delta^\beta(\omega)$ with $\beta \in \{ R,K \}$, is given by
\begin{equation}
 \label{eq:tot_hyb}
 \Delta^\beta(\omega) = \sum_{\lambda} \Delta^\beta_\lambda(\omega)\,.
\end{equation}
Notice that $\Delta^\beta(\omega)$ encodes the combined effect of $H_{\mathrm{leads}}$ and $H_{\mathrm{coup}}$ on the impurity.
Thus, the properties of the impurity are controlled by $\Delta^\beta(\omega)$ and by $H_{\mathrm{imp}}$, alone.
\subsection{Nonequilibrium Green's function}
\label{subsec:neq_gfs}
Out of equilibrium, there are two independent single-particle Green's functions.
We are especially interested in the steady state Green's functions at the impurity.  
The lesser and the greater one are defined as,
\begin{equation}
\label{eq:Glesser_Ggreater}
\begin{split}
  G^<_\sigma(t) &= \hphantom{-}i \Braket{f_\sigma^\dagger(t) f_{\sigma}^\nag}_\infty \,,\\
  G^>_\sigma(t) &= -i \Braket{f_\sigma^\nag(t) f_{\sigma}^\dagger}_\infty \,.
\end{split}
\end{equation}
Note that they have only one time argument, since in steady state (indicated by the subscript $\infty$), the system is time-translation invariant.
After a Fourier transform to frequency space, 
\begin{equation}
\label{eq:FTGF}
 G^\alpha_\sigma(\omega) = \int G^\alpha_\sigma(t) \exp(i\omega t) \,dt \,,
\end{equation}
with $\alpha \in \{<,>\}$, these Green's functions may be combined to obtain the spectral function or local impurity DOS and the Keldysh Green's function, which we are typically interested in, 
\begin{align}
  A_\sigma(\omega) &= \frac{i}{2 \pi} \left[ G^>_\sigma(\omega) - G^<_\sigma(\omega)\right] \label{eq:A_from_G} \,,\\
  G^K_\sigma(\omega) &= G^>_\sigma(\omega) + G^<_\sigma(\omega) \label{eq:GK_from_G} \,.
\end{align}
From the spectral function the retarded and the advanced Green's function are obtained via the Kramer's Kronig relations.

In the nonequilibrium Green's function formalism $G^R_\sigma(\omega)$, $G^A_\sigma(\omega)$ and $G^K_\sigma(\omega)$ 
are typically arranged in a $2 \times 2$ matrix (Keldysh space), which we indicate by an underline,  
\begin{equation*}
 \underline{G}_\sigma(\omega) \equiv \begin{pmatrix} G^R_\sigma(\omega) & G^K_\sigma(\omega) \\ 0 & G^A_\sigma(\omega) \end{pmatrix} \,.
\end{equation*}
This has the advantage that Dyson's equation is valid in the same form as in equilibrium,
\begin{equation}
\label{eq:total_Dyson}
\begin{split}
 \underline{G}^{-1}_\sigma(\omega) &= \underline{G}^{-1}_{0 \sigma}(\omega) -\underline{\Sigma}(\omega) \,,\\
 \underline{G}^{-1}_{0 \sigma}(\omega) &=\underline{g}^{-1}_{0 \sigma}(\omega) - \underline{\Delta}(\omega) \,.
\end{split}
\end{equation}
Here, $\underline{g}_{0 \sigma}$ is the Green's function of the decoupled and noninteracting impurity, the self energy $\underline{\Sigma}(\omega)$ accounts for the interaction, and the hybridization function $\underline{\Delta}(\omega)$ for the coupling to the noninteracting leads. 

From the Green's functions defined above, the current across the impurity can be obtained as
\begin{equation}
 j_\lambda = \frac{1}{2\pi} \sum_{\sigma} \int \rre{\left( G^R_\sigma \Delta^K_\lambda + G^K_\sigma \Delta^R_\lambda \right)} \, d\omega \,.\label{eq:curr_formula}
\end{equation}
In steady state, the left and right-moving current must be identical, $|j_L| = |j_R|$, so we can also compute $j = \frac{1}{2} \left( j_R - j_L \right)$. 
The differential conductance follows from the current via
\begin{equation}
 G = \frac{dj}{d\phi} \label{eq:G_general_formula} \,.
\end{equation} 
\subsection{Auxiliary master equation approach}
\label{sec:amea}
The auxiliary master equation approach (AMEA) is based upon a mapping of the model introduced in \se \ref{subsec:model}  --
which we call {\em physical} system in the following -- consisting of an impurity and an infinite bath,  to a finite {\em auxiliary} open quantum system.
The latter consists of the impurity coupled to a small number of $N_B=N-1$ auxiliary bath sites that are furthermore attached to Markovian environments. 
The dynamics of the auxiliary system is governed by a Lindblad master equation,\cite{do.nu.14} whose parameters are chosen such that its hybridization function $\underline{\Delta}_{\mathrm{aux}} $ approximates the one of the physical system $\underline{\Delta}_{\mathrm{phys}}$ (\eq{eq:tot_hyb}) as accurately as possible. 
Upon solving the corresponding many-body Lindblad equation, an approximation for the behavior of the interacting impurity in the physical system is found.
We stress that this mapping becomes exponentially exact, upon increasing the number of bath sites $N_B \to \infty$ in the sense that the Lindblad bath provides an exponentially accurate representation of the original Hamiltonian problem.\cite{do.so.17,ch.ar.19} 

\subsubsection{Lindblad equation}
\label{subsubsec:Lindblad}
As outlined in \tcites{do.nu.14,dz.ko.11}, the Lindblad equation for a fermionic lattice model can be expressed in terms of an ordinary Schr\"odinger equation in an augmented state space of twice as many sites $2N$, 
\begin{equation}
 \label{eq:Lindblad_as_Schroedinger}
 \frac{d}{dt}\ket{\rho(t)} = L \ket{\rho(t)} \,.     
\end{equation}
In this augmented space, the density operator is represented by a quantum state $\ket{\rho(t)}$ and the Lindbladian $i L$ plays the role of a non-Hermitian Hamiltonian. 
For our case,\footnote{
See, e.g., Eqs.~(9)-(11) in \tcite{do.nu.14}.
} 
it reads
\begin{equation}
\label{eq:iL}
\begin{split}
 i L &= \sum_\sigma \boldsymbol{c}_\sigma^\dagger \!
 \begin{pmatrix}
  \boldsymbol{E} + i\boldsymbol{\Omega} & 2 \boldsymbol{\Gamma}^{(2)} \\
  -2 \boldsymbol{\Gamma}^{(1)} & \boldsymbol{E} - i\boldsymbol{\Omega}
 \end{pmatrix}
 \!\boldsymbol{c}_\sigma^\nag -2 \, \mathrm{Tr} \left( \boldsymbol{E} +i \boldsymbol{\Lambda} \right) \\
 &\hphantom{=} + U \left( n_{f\uparrow}^\nag n_{f\downarrow}^\nag - \tilde{n}_{f\uparrow}^\nag \tilde{n}_{f\downarrow}^\nag + \sum_{\sigma} \tilde{n}_{f\sigma}^\nag -1 \right) \,.\\
\end{split}
\end{equation}
Here, $\boldsymbol{E}$, $\boldsymbol{\Gamma}^{(1)}$ and $\boldsymbol{\Gamma}^{(2)}$ are $N \times N$ matrices holding the parameters of the Lindblad equation yet to be determined by a fit of $\underline{\Delta}_{\mathrm{aux}}$ to $\underline{\Delta}_{\mathrm{phys}}$ and
\begin{equation}
\begin{split}
 \label{eq:Omega_Lambda}
 \boldsymbol{\Omega} &= \boldsymbol{\Gamma}^{(2)} - \boldsymbol{\Gamma}^{(1)} \,,\\
 \boldsymbol{\Lambda} &= \boldsymbol{\Gamma}^{(2)} + \boldsymbol{\Gamma}^{(1)} \,.
\end{split}
\end{equation}
The vector
\begin{equation}
 \label{eq:c_vector}
 \boldsymbol{c}_\sigma^\dagger = \left( c_{1\sigma}^\dagger , \ldots , c_{N\sigma}^\dagger , \tilde{c}_{1\sigma}^\dagger , \ldots , \tilde{c}_{N\sigma}^\dagger \right) \,
\end{equation}
contains the creation operators $c_{i\sigma}^\dagger$ and $\tilde{c}_{i\sigma}^\dagger$ in the auxiliary system, which is composed of original~\footnote{
``original'' refers to Eqs.~(9)-(11) in \tcite{do.nu.14}.
} 
\nontildedef and additional ``tilde'' sites.
They obey the usual fermionic anticommutation rules.  
$f$ is the position of the impurity site, which is typically in the center, $f=(N+1)/2$, $n_{f\sigma} \equiv c_{f\sigma}^\dagger c_{f\sigma}^\nag$ and $\tilde{n}_{f\sigma}$ analogously.

In this framework, steady state expectation values as well as Green's functions are obtained as~\footnote{
Here, we have used the fact that $\bra{I}L=0$.
}
\begin{equation}
 \braket{A(t)B} = \braket{I|Ae^{L t}B|\rho_{\infty}} \,,
\end{equation}
for local impurity operators $A,B$ and times $t \geq 0$.
Here, $\ket{\rho_{\infty}} = \lim_{t \to \infty} \ket{\rho(t)}$ defines the steady state %
and $\ket{I}$ is the so-called left vacuum,\footnote{
This representation of $\ket{I}$ follows from Eq.~(13) in \tcite{dz.ko.11} via particle-hole transformation.
}
\begin{align}
 \label{eq:Ivac}
 \ket{I} &= \sum_{\{ \underline{n} \}} \ket{\underline{n},\tilde{\underline{n}}} \,, \\
 \ket{\underline{n},\tilde{\underline{n}}} &\equiv (-i)^{\sum_{i \sigma} n_{i \sigma}} ( c_{1 \sigma}^\dagger \tilde{c}_{1 \sigma}^\nag )^{n_{1 \sigma}} \sdots ( c_{N \sigma}^\dagger \tilde{c}_{N \sigma}^\nag )^{n_{N \sigma}} \ket{0}\!\ket{\tilde{F}} \,. \nonumber
\end{align}
$n_{i\sigma}$ and $\ket{0}$ are the occupation numbers and the vacuum in the \nontilde system and $\ket{\tilde{F}}$ is the completely filled Fock state in the tilde system.
\eqs{eq:Lindblad_as_Schroedinger}-\eqref{eq:Ivac} describe the so-called super-fermion (SF) representation. 
\subsubsection{Mapping procedure}
\label{subsubsec:Mapping}

The mapping to the auxiliary system is outlined in \tcites{do.so.17, so.fu.19} and we sketch it only briefly, here. 
Starting from proper initial values, the parameters $\boldsymbol{E}_{ij}, \boldsymbol{\Gamma}_{ij}^{(1)}, \boldsymbol{\Gamma}_{ij}^{(2)}$ are adjusted by minimizing a suitable~\cite{do.nu.14,do.so.17} cost function. 
This cost function punishes deviations between the auxiliary and the physical hybridization function and, in general, both the retarded and Keldysh component contribute. 
Its evaluation involves only the solution of a noninteracting problem, which is computationally cheap. 
In this paper, the optimizaton  of the Lindblad parameters is carried out with the ADAM~\cite{ki.ba.14u}  algorithm as implemented in the python library tensorflow.\cite{ab.ag.16u}

In principle, the best fit is obtained by allowing the Lindblad parameters to connect all pairs of lattice sites.\cite{do.so.17} 
However, employing matrix product states (MPS) as solver for the many-body problem, as described in \se \ref{subsubsec:MPS}, it is convenient to adopt a one-dimensional geometry, which minimizes the entanglement. 
Specifically, here we adopt a chain geometry with the impurity in the center.
In this case, the optimal solution numerically turns out to be such that all sites to the left (right) of the impurity have 
$\boldsymbol{\Gamma}^{(2)} = 0$ ($\boldsymbol{\Gamma}^{(1)} = 0$) and, therefore, are almost completely \almostempty (\almostfull).\cite{do.ga.15}
This situation is particularly convenient for the MPS many-body solution, since it prevents the propagation of entanglement, as discussed in \tcite{do.ga.15}.
In addition, knowing this fact, it is then sufficient to fit the retarded component of the hybridization function, only, as explained in \app \ref{subsec:F_E_bath}.

We start from the zero-bias, $\phi=0$, i.e. equilibrium situation and perform the fit as discussed above.
The important physics obviously occurs in the region around $\omega = 0$ and is controlled by the power-law exponent $r$. 
Thus, it is particularly important to have an accurate fit there.
In order to achieve this, we introduce a weight in the cost function, which is twice as large on $|\omega|\leq 1$ than on $|\omega|>1$.  
For nonzero $\phi$, we can construct the nonequilibrium fit from the equilibrium one, as outlined in \app \ref{subsec:EQ_bath_to_NEQ_bath}.
This has the advantage that the accuracy of the fit to reproduce the power-law is independent of the bias voltage, which is crucial, in order to faithfully investigate the crossover to finite voltage.
\subsubsection{Matrix product states implementation}
\label{subsubsec:MPS}
We solve the many-body Lindblad equation employing matrix product states (MPS) in combination with the time-dependent density matrix renormalization group (tDMRG) algorithm.\cite{wh.fe.04,da.ko.04}
MPS are especially suited for one-dimensional problems, where they can provide an efficient representation with a small bond dimension. 
In particular, ground states of one-dimensional gapped closed systems are conveniently expressed as MPS.\cite{scho.11}
On the other hand, also steady states and Green's functions of open quantum systems in a chain geometry are reproduced accurately using MPS and the entanglement remains limited.\cite{do.ga.15}
We decided to employ tDMRG for the time evolution here, since it is conveniently implemented with the C++ tensor network library iTensor.\cite{ITensor}

\begin{center}
\begin{figure}[h]
  \begin{minipage}[b]{0.9\columnwidth} %
    \centering \includegraphics[width=\textwidth]{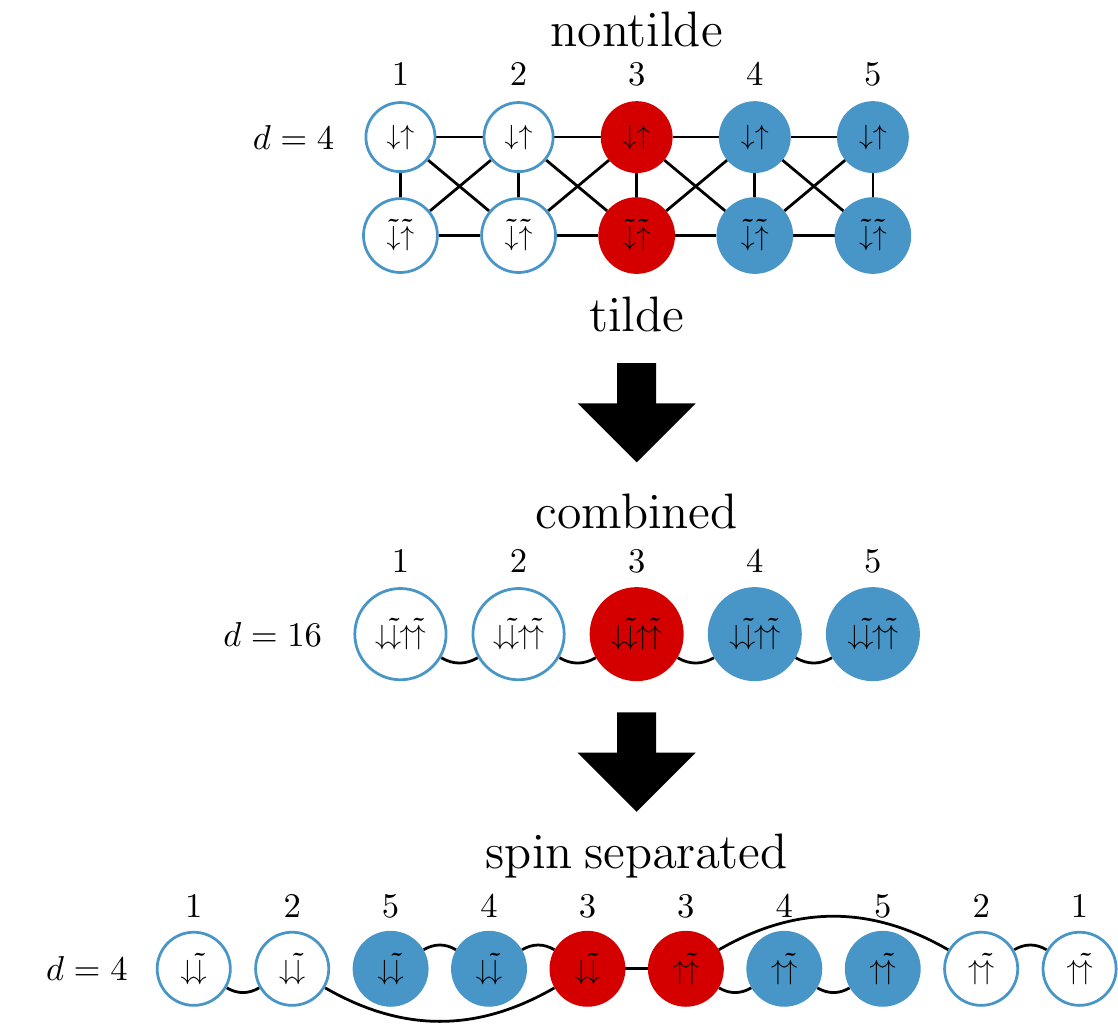} 
  \end{minipage}
\caption{
Construction of effective sites for the MPS time evolution.
The impurity sites are displayed as red circles, the  \almostfull and \almostempty bath sites as blue and white ones.
As discussed in the text, by \almostfulldef and \almostemptydef we mean sites for which $\boldsymbol{\Gamma}^{(1)} = 0$ or $\boldsymbol{\Gamma}^{(2)} = 0$, respectively, for details, see \app \ref{subsec:F_E_bath}.
Each site is labelled with an index and its spin and tilde degrees of freedom. 
The upper panel of this figure shows the sites and their couplings occurring in the Lindblad equation in the augmented state space. 
Here, the upper (lower) part of this ladder structure is formed by \nontilde (tilde) sites. 
Lines connecting these two sets of sites represent $\Gamma$ terms, while lines within the same set are hoppings. 
The central panel shows the effective sites used in \tcite{do.ga.15} that result from combining \nontilde and tilde sites with the same index. 
Finally, the lower panel shows the effective sites used in this work that result from the combined sites by separating the spin degrees of freedom. The advantage of this representation is that the local Hilbert space has a dimension of $4$, instead of $16$ as in our previous work. On the other hand, it introduces two long-range hopping terms.
}
\label{fig:sketch_effective_sites}
\end{figure}
\end{center}

Within AMEA, a chain geometry naturally results from combining a \nontilde and a tilde site associated with an index $i$, according to \eq{eq:c_vector}, to a single effective site with a local Hilbert space dimension of $d=16$,\cite{do.ga.15} see \fig{fig:sketch_effective_sites}.
Since the SIAM couples opposite spins only at the impurity, it is convenient to separate spin-up and spin-down degrees of freedom,\cite{ba.zi.17} which reduces the local Hilbert space dimension back to $d=4$.
\fig{fig:sketch_effective_sites} shows the effective sites we use in this work (lower panel) and sketches the steps to obtain them. 
Note that in this arrangement, the Hubbard interaction is on the bond between the spin-down and spin-up impurity site.
Furthermore, it is necessary to introduce two long-range terms between the \almostempty bath sites and the impurity, violating the linear geometry.

We encode the left vacuum $\ket{I}$ as well as a proper initial state $\ket{\rho(t=0)}$ as MPS on these effective sites.  
We choose $\ket{\rho(0)} \propto \ket{I}$, since this has proved convenient in our previous work.\cite{do.ga.15,fu.do.18}
Taking
\begin{equation}
\begin{split}
 &\ket{n_{1 \downarrow} \tilde{{n}}_{1 \downarrow} \ldots n_{f-1 \downarrow} \tilde{{n}}_{f-1 \downarrow} n_{N \downarrow} \tilde{{n}}_{N \downarrow} \ldots n_{f \downarrow} \tilde{{n}}_{f \downarrow} }  
 \\ 
 \otimes& \ket{n_{f \uparrow} \tilde{{n}}_{f \uparrow} \ldots n_{N \uparrow} \tilde{{n}}_{N \uparrow} n_{f-1 \uparrow} \tilde{{n}}_{f-1 \uparrow} \ldots n_{1 \uparrow} \tilde{{n}}_{1 \uparrow} }   
 \end{split}
\end{equation}
as basis states, we can express the corresponding expansion coefficients $\psi(\{ n_{i \sigma}, \tilde{n}_{i \sigma} \})$ of any required state as products of local matrices,
\begin{equation}
 \label{eq:MPS}
 \begin{split}
\psi(\{ n_{i \sigma}, \tilde{n}_{i \sigma} \})
 =&\, \boldsymbol{A}^{n_{1 \downarrow} \tilde{n}_{1 \downarrow}} \ldots \boldsymbol{A}^{n_{f-1 \downarrow} \tilde{n}_{f-1 \downarrow}} 
 \\
 \times&\, \boldsymbol{A}^{n_{N \downarrow} \tilde{n}_{N \downarrow}} \ldots \boldsymbol{A}^{n_{f \downarrow} \tilde{n}_{f \downarrow}} 
 \\
 \times&\, \boldsymbol{A}^{n_{f \uparrow} \tilde{n}_{f \uparrow}} \ldots \boldsymbol{A}^{n_{N \uparrow} \tilde{n}_{N \uparrow}} 
 \\
 \times&\, \boldsymbol{A}^{n_{f-1 \uparrow} \tilde{n}_{f-1 \uparrow}} \ldots \boldsymbol{A}^{n_{1 \uparrow} \tilde{n}_{1 \uparrow}} \,.
 \end{split}
\end{equation}
In case of $\ket{I}$, only matrices with $n_{i \sigma} = 1 - \tilde{n}_{i \sigma}$ are nonzero.
Specifically, comparing with \eq{eq:Ivac}, the corresponding expansion coefficients read
\begin{equation}
 \psi(\{ n_{i \sigma}, \tilde{n}_{i \sigma} \}) = \prod_{i\sigma} \delta_{n_{i \sigma}, 1 - \tilde{n}_{i \sigma}} (-i)^{n_{i\sigma}} \,,
\end{equation}
resulting in the $1\times 1$, i.e. scalar matrices $\boldsymbol{A}^{01} = 1$ and $\boldsymbol{A}^{10} = -i$. 
Having expressed the relevant states as MPS, we can proceed with the time evolution of the auxiliary system.

In tDMRG the time evolution of the system, $\ket{\rho(t)} = \exp{(Lt)} \ket{\rho(0)}$, is decomposed into a Trotter sequence of small time evolutions on bonds induced by gates.
After the application of a gate, the original structure of the MPS, \eq{eq:MPS}, is restored with a singular value decomposition. 
As usual at this step, the smallest singular values are neglected defining a truncated weight, which is the sum of all discarded squared singular vales.
Then the next gate may be applied in the same way.\cite{scho.11}

\begin{center}
\begin{figure}[h]
  \begin{minipage}[b]{1.3\columnwidth} %
    \centering \includegraphics[width=\textwidth]{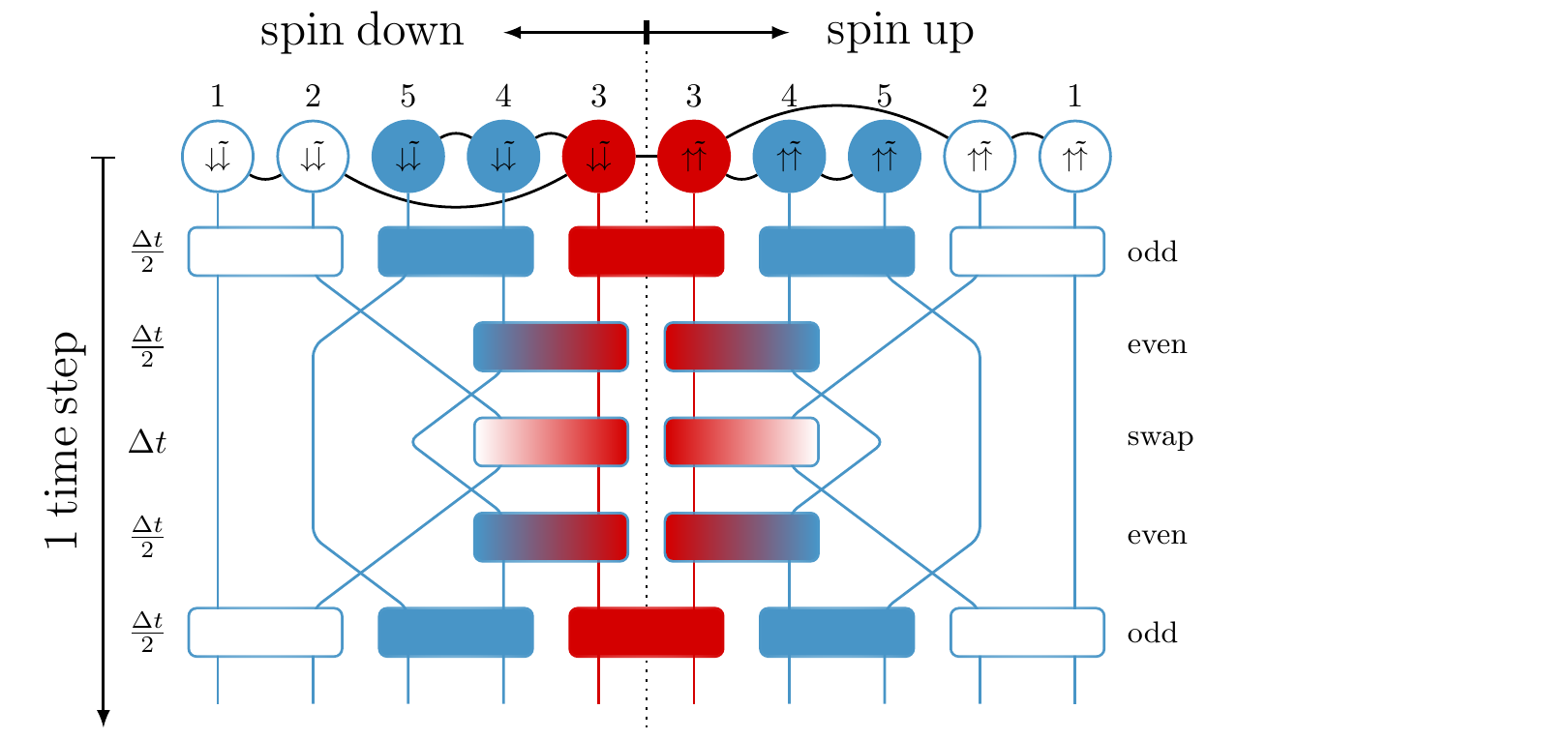} 
  \end{minipage}
\caption{
Single step in the MPS time evolution of the (PSG) SIAM with separated spin degrees of freedom. 
The impurity sites are represented as red circles, the \almostfull and \almostempty bath sites as blue and white ones.
The same colouring also classifies the time evolution gates that are represented as boxes.
A time evolution step $\Delta t$ consists of five layers, labelled ``odd'', ``even'' and ``swap''.
In each layer, a site $i\sigma$, with index $i$ and spin $\sigma$, is touched only by one gate.
In the swap layer, swap gates displayed as crossing time lines are employed to cope with the long-range couplings between the \almostempty bath sites and the impurity sites.
}
\label{fig:sketch_tensor_network}
\end{figure}
\end{center}

\fig{fig:sketch_tensor_network} shows the sequence of gates we use in this work to evolve one time step $\Delta t$. 
There are five layers, labelled ``odd'', ``even'' and ``swap'', and the gates within them are displayed as boxes.
In order to understand them, we identify the following terms as building blocks of the Lindbladian, \eq{eq:iL}, 

\begin{equation}
\label{eq:iL_building_blocks}
\begin{split}
 i L_{i\sigma j\sigma} &= \left(\boldsymbol{E} + i\boldsymbol{\Omega}\right)_{ij} c^\dagger_{i\sigma} c^\nag_{j\sigma}  -2 \boldsymbol{\Gamma}^{(1)}_{ij} \tilde{c}^\dagger_{i\sigma} c^\nag_{j\sigma}
 \\
 &+ 2 \boldsymbol{\Gamma}^{(2)}_{ij} c^\dagger_{i\sigma} \tilde{c}^\nag_{j\sigma} + \left(\boldsymbol{E} - i\boldsymbol{\Omega}\right)_{ij} \tilde{c}^\dagger_{i\sigma} \tilde{c}^\nag_{j\sigma} \,,
 \\
 i L_{f\uparrow f\downarrow} &= U \big( n_{f\uparrow}^\nag n_{f\downarrow}^\nag - \tilde{n}_{f\uparrow}^\nag \tilde{n}_{f\downarrow}^\nag + \tilde{n}_{f\uparrow}^\nag + \tilde{n}_{f\downarrow}^\nag \big) \,.
\end{split}
\end{equation}

Within the odd layers, all on-site terms in \eq{eq:iL_building_blocks} as well as the two-site terms on every second bond, according to \fig{fig:sketch_tensor_network}, including the impurity bond, are grouped, exponentiated and applied as gates, see \eq{eq:iL_gates}. 
In the even layers, the two-site gates on the remaining bonds are applied, excluding the long-range bonds between the impurity and the \almostempty baths, which are taken care of in the swap layer.\cite{scho.11,st.wh.10}
In the swap layer, the innermost sites of the \almostempty baths are swapped with their nearest neighbors, i.e. they change positions, until they are next to the impurity sites.
Then the time evolution gates are applied, before they are swapped back to their original positions. 
Swap gates are displayed as crossing time lines.
Summarizing:
\begin{equation}
\label{eq:iL_gates}
\begin{split}
\mathrm{odd:} & \quad
 \begin{cases}
  &\,\exp{\left[ (L_{i\sigma j\sigma}+L_{j\sigma i\sigma}+L_{i\sigma i\sigma}+L_{j\sigma j\sigma}) \tfrac{\Delta t}{2} \right]} ,\\ 
  &\quad\quad (i,j)=\{(1,2),(4,5),\cdots\} \\\\
  &\,\exp{\left[ (L_{f\uparrow f\downarrow}+L_{f\uparrow f\uparrow}+L_{f\downarrow f\downarrow}) \tfrac{\Delta t}{2} \right]}
 \end{cases} 
\\[5pt]
\mathrm{even:}  &\quad\quad\quad \exp{\left[ (L_{i\sigma j\sigma}+L_{j\sigma i\sigma}) \tfrac{\Delta t}{2} \right]} ,
\\ 
&\quad\quad\quad\quad\quad (i,j)=\{(3,4),\cdots\} 
\\[5pt]
\mathrm{swap:}  &\quad\quad\quad \exp{\left[ (L_{f-1\sigma f\sigma}+L_{f\sigma f-1\sigma}) \, \Delta t \, \right]} 
\end{split}
\end{equation}
To complete the time step, also the constant in \eq{eq:iL} has to be taken into account, so we multiply the MPS with $\exp{\{ i \Delta t \, [ 2 \mathrm{Tr} \left( \boldsymbol{E} +i \boldsymbol{\Lambda} \right) + U] \} }$.

Notice that the described sequence of gates may be employed, provided that $N_B$ is even, as reasonable at PH symmetry, otherwise the sequence needs to be adjusted accordingly.
Since this sequence is derived from a second-order Suzuki-Trotter decomposition, an error $\mathcal{O}(\Delta t^3)$ is acquired in the time evolution, which is further proportional to the commutators of the Lindbladians, \eq{eq:iL_gates}, in different layers.
Additionally, there is an error from the truncation of the singular values after the application of each gate.

In this work, we employ the tDMRG scheme as follows:
We first determine the steady state $\ket{\rho_\infty} \propto \exp(L t^*) \ket{I}$~\footnote{
$\braket{I|\rho(t)} = 1$ must be fulfilled for all $t$, since this corresponds to $\Tr{\rho(t)} = 1$.
}
via time evolution of the initial state with tDMRG up to a time $t^*$, for which expectation values of static observables, such as single and double occupancies, are converged.
Afterwards, we compute, e.g., the lesser impurity Green's function, $G^<_\sigma(t) = i \braket{I|c_{f\sigma}^\dagger \exp(L t) c_{f\sigma} |\rho_\infty}$, by applying $c_{f\sigma}$ to the steady state, employing tDMRG again, applying $c_{f\sigma}$ to $\ket{I}$ and calculating the overlap.  
$G^<_\sigma(\omega)$ is obtained in the frequency domain via Fourier transformation of $G^<_\sigma(t)$ after linear prediction.\cite{ba.sc.09}
\subsubsection{Physical versus auxiliary quantities}
\label{subsubsec:phys_vs_aux}
The observables obtained directly by the MPS treatment of the auxiliary system are called ``auxiliary'' quantities in the following.
The auxiliary Green's functions are used as an approximation for the  Green's functions of the physical model. 
As discussed, this approximation becomes exponentially exact upon increasing the number of bath sites.
We can get an even better approximation by extracting the self energy from Dyson's equation for the auxiliary system, assuming $\underline{\Sigma}_{\mathrm{phys}}(\omega) \approx \underline{\Sigma}_{\mathrm{aux}}(\omega)$ and reentering Dyson's equation with the (approximated) physical self energy and the (exact) physical hybridization function.
The Green's functions extracted in this way are refereed to as ``physical'' in the following. 
\section{Results}
\label{sec:results}
Here, we present results obtained with AMEA for the parameters $r=0.25$, $U=6$, $T = 0.05$ and $\Gamma = 1$ in the generalized Kondo (GK) phase and $\Gamma = 0.25$ in the local moment (LM) phase. 
In equilibrium, we compare the results with the ones obtained with a direct MPS time evolution of the Hamiltonian, \eq{h}, at $T=0$.\cite{ba.zi.17} 
For clarity, we refer to this procedure as ``Hamiltonian MPS'' (HMPS), in order to distinguish it from AMEA, which is also treated via MPS. 
Of course, HMPS cannot be used to achieve the steady state, since the system is finite.
Since HMPS is faster, we also provide equilibrium results for different values of $r$ and $U$ obtained with that approach.
\subsection{Fit}
\label{subsec:results_mapping}
\begin{figure}[h!]
\subfigure[]{
  \label{subfig:DeltaR_eq}
  \begin{minipage}[b]{\pw\columnwidth}
    \centering \includegraphics[width=1\textwidth]{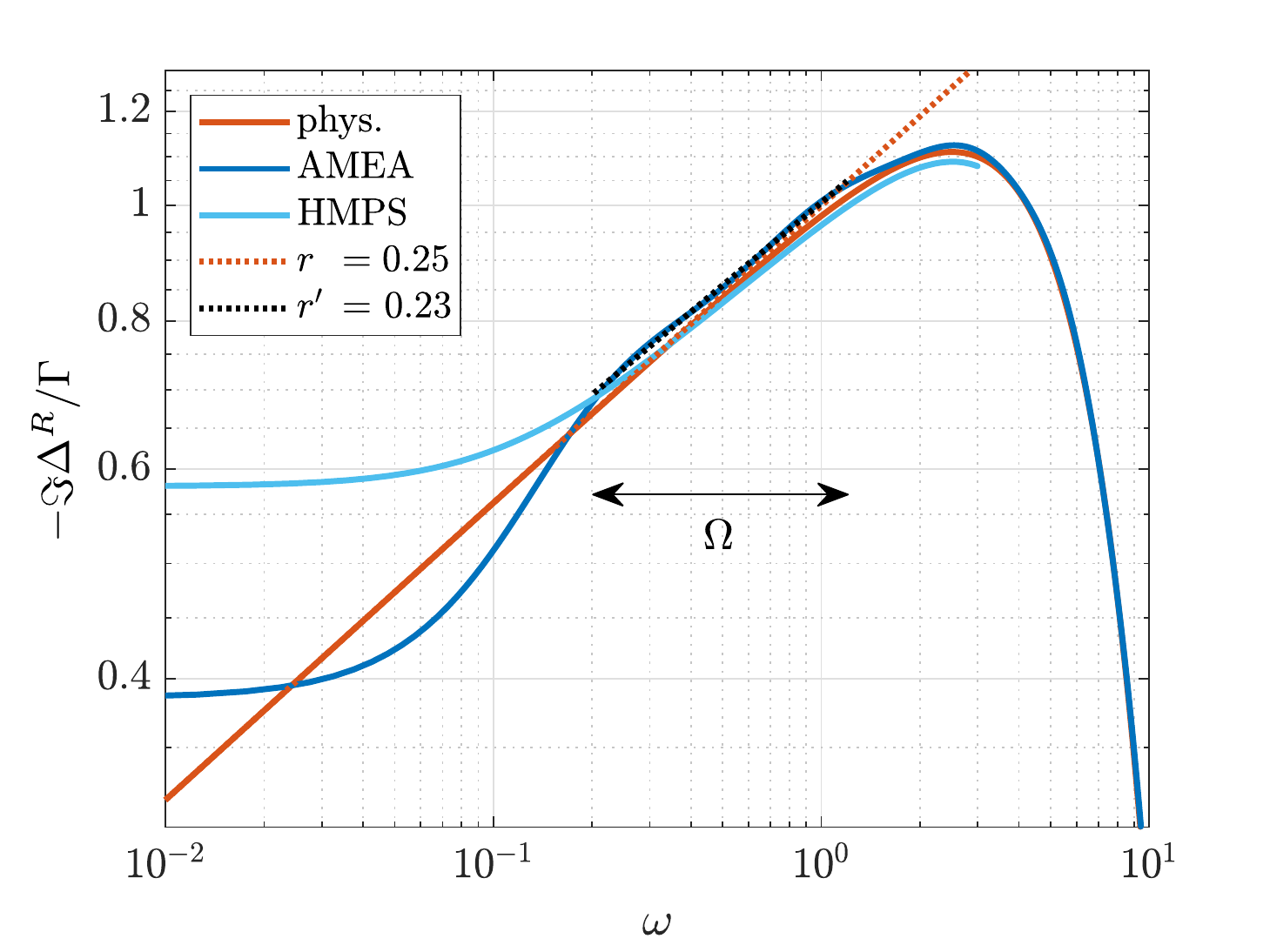}
  \end{minipage}
}\\
\subfigure[]{
  \label{subfig:distfun_eq}
  \begin{minipage}[b]{\pw\columnwidth}
    \centering \includegraphics[width=1\textwidth]{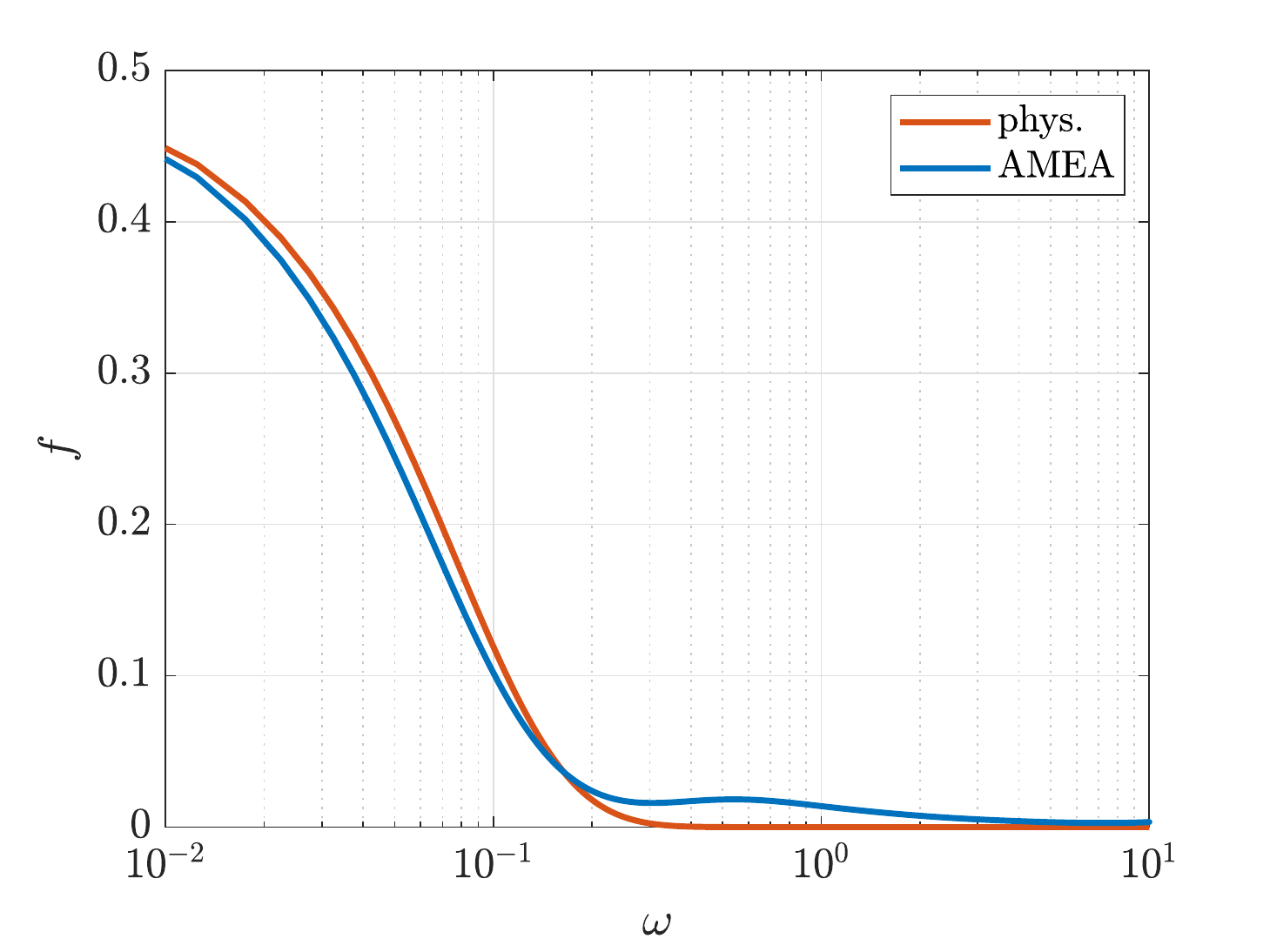} %
  \end{minipage}
}
\caption{
Equilibrium ($\phi = 0$) fit results.
\subref{subfig:DeltaR_eq} Retarded hybridization function $-\Im \Delta^R(\omega)$ in units of the hybridization strength $\Gamma$ and \subref{subfig:distfun_eq} distribution function $f$ determined from $\Im \Delta^R$ and $\Im \Delta^K$ via \eq{eq:hyb_K}. 
The power-law exponent $r^\prime$ is obtained by fitting the AMEA hybridization function with \eq{eq:PL_Delta} on the interval $\Omega$. 
The same procedure applied to the HMPS result yields quite the same exponent (up to a deviation of $\approx 0.01$).
$|\omega|^r$ is plotted for comparison, see \eq{eq:PL_Delta}.
These curves are hardly distinguishable (black vs. red dots).
}
\label{fig:eq_fit}
\end{figure}

We start by fitting the equilibrium hybridization function with the auxiliary Lindblad system, as described in \se \ref{subsubsec:Mapping}. 
As discussed above, we use a weight function, such that the hybridization function is reproduced better at low frequencies. 
We also concentrate on reproducing the power-law as accurately as possible, while putting less emphasis on the  multiplicative factors as well as on the large-$\omega$ behavior.
The results of the fit are displayed in \fig{fig:eq_fit}.

From \fig{subfig:DeltaR_eq} we can see that the auxiliary (AMEA) retarded hybridization function accurately matches the physical one for $|\omega| \gtrsim 0.2$, which, on the other hand, behaves approximately as  
\begin{equation}
 \label{eq:PL_Delta}
 \Im \Delta^R(\omega) \propto |\omega|^r 
\end{equation}
for $|\omega| \lesssim 1.2$.
It follows that $\Im \Delta^R_\mathrm{aux}(\omega)$ displays a power-law on the interval $\Omega \equiv (0.2 < |\omega| < 1.2)$, but the exponent is slightly underestimated.
In fact, a fit by Eq.~\ref{eq:PL_Delta} on the interval $\Omega$ yields $r^\prime = 0.23$, whereas its value should be equal to $r = 0.25$.
Note that the behavior of $-\Im \Delta^R_\mathrm{aux}(\omega)$ is qualitatively acceptable~\footnote{
in the sense that it is decreasing
}
even down to $|\omega| \approx 0.02$, which is one order of magnitude smaller than the lower edge of the power-law interval $\Omega$. 
Below this value, though, it bends towards a constant, $-\Im \Delta^R_\mathrm{aux}(\omega=0) \approx 0.39 \Gamma$, instead of going to zero.
\fig{subfig:DeltaR_eq} also shows the hybridization function used in HMPS, for comparison. 
Here, it is plotted using a Lorentzian broadening of $\eta = 0.1$.\footnote{
In HMPS, in the Fourier transform of the Green's function, \eq{eq:FTGF}, a modified kernel $\exp{(i \omega t - \eta |t|)}$ with a finite broadening $\eta$ is used, instead of the mathematically exact limit $\eta \to 0$. 
Note that in $\omega$-space this is equivalent to a convolution of the exact (finite size) result with a Lorentzian distribution of width $\eta$.
Here, $\eta$ is chosen such that the hybridization function, the spectral function and the self energy are smooth.
}
It features a good representation of the power-law, roughly on the same interval $\Omega$ as AMEA, but $-\Im \Delta^R_\mathrm{HMPS}(0)$ is larger for this value of $\eta$. 
Note that for HMPS many more bath sites are necessary to get such a high resolution. 
Specifically, on $|\omega|<10$ we use $N_B=1301$ for HMPS in comparison to $N_B = 10$ or $20$ for AMEA~\footnote{
$N_B = 20$ in nonequilibrium, see \app \ref{sec:NEQ_system_construction}
} 
to achieve roughly the same accuracy.
In \fig{subfig:distfun_eq} the auxiliary distribution function $f_\mathrm{aux}$, obtained from $\Delta^R_\mathrm{aux}(\omega)$ and $\Delta^K_\mathrm{aux}(\omega)$ via \eq{eq:hyb_K}, is plotted. 
It compares well to the Fermi function, i.e. the distribution function in the physical system.

Since $\Omega$ identifies the interval, where we can faithfully represent the power-law in AMEA and in HMPS with an exponent $r^\prime \approx r$, it is also the interval, where we should study other quantities, such as the spectral function $A(\omega)$ or the self energy $\Sigma^R(\omega)$. 
With a bias voltage applied, the interval $\Omega$ shrinks to
\begin{equation}
\label{eq:Omega_phi_int}
 \Omega(\phi)= \left(0.2 + \frac{\phi}{2} < |\omega| < 1.2 - \frac{\phi}{2} \right), \quad \mathrm{for}~\phi\geq0 \,, 
\end{equation}
since the hybridization functions $\Delta_L^R$ and $\Delta_R^R$ are shifted by $\phi$ with respect to each other. 
This also limits the values of the bias voltage, in which we can reasonably estimate a power-law behavior to $\phi \lesssim 0.6$. 
This estimate is obtained by assuming that we need a frequency interval of width $\epsilon = 0.4$, in which to fit power-law exponents. 
\subsection{Many-body solution}
After carrying out the fit, we solve the resulting Lindblad equation (or Schr\"odinger equation in case of HMPS) and determine the steady state (or just equilibrium for HMPS) Green's functions, as described in \se \ref{subsubsec:MPS} (or \tcite{ba.zi.17}).
We are especially interested in the spectral function as well as the self energy, as there are predictions about their behavior in equilibrium,\cite{gl.lo.00} and in the differential conductance.
Unless stated otherwise, our plots display the physical spectral functions and not the auxiliary ones, acoording to the definition in \se \ref{subsubsec:phys_vs_aux}. 
Due to the Trotter and truncation errors, the MPS results break PH symmetry. 
Therefore, the curves we show are PH symmetrized and the shadings indicate an estimate of these errors obtained from the deviations from PH symmetry, see \app \ref{sec:error_estimation} for a more detailed discussion.
\subsubsection{Equilibrium case}
\label{subsubsec:equilibrium}
\begin{figure}[h]
\subfigure[]{
  \label{subfig:A_eq}
  \begin{minipage}[b]{\pw\columnwidth}
    \centering \includegraphics[width=1\textwidth]{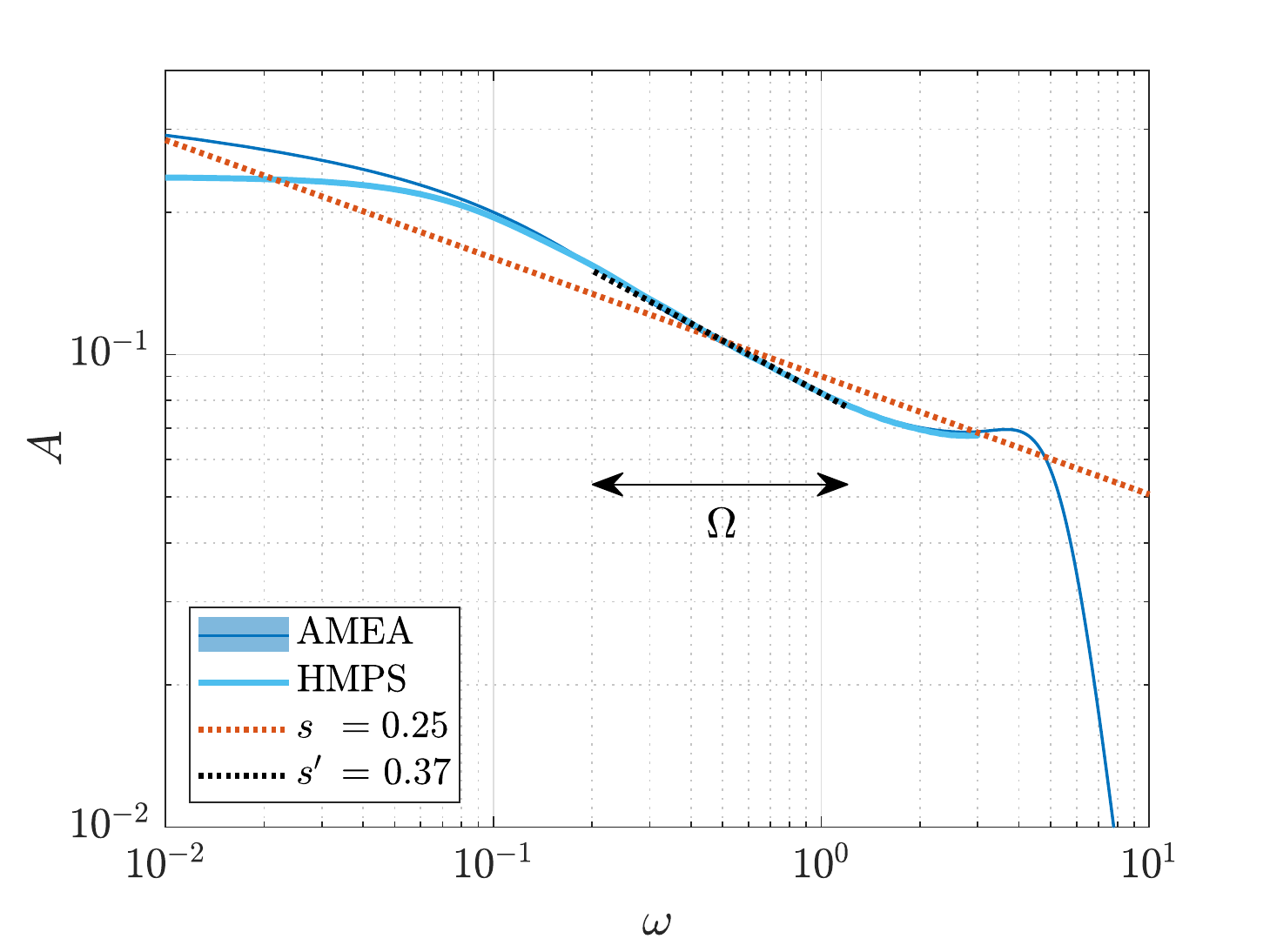} 
  \end{minipage}
}\\
\subfigure[]{
  \label{subfig:SigmaR_eq}
  \begin{minipage}[b]{\pw\columnwidth}
    \centering \includegraphics[width=1\textwidth]{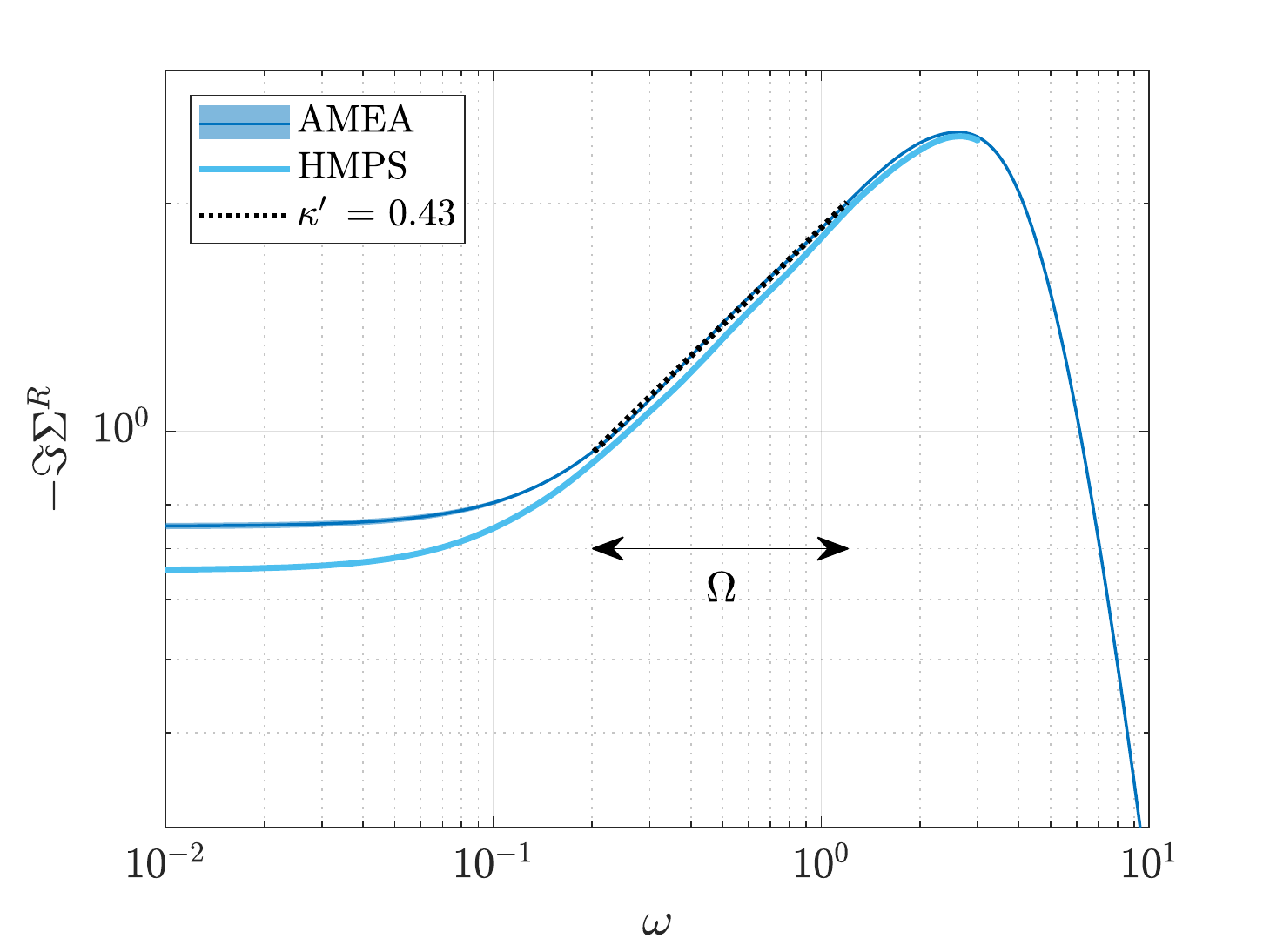} 
  \end{minipage}
}\\
\caption{
Equilibrium ($\phi = 0$) \subref{subfig:A_eq} spectral function $A(\omega)$ and \subref{subfig:SigmaR_eq} retarded self energy $-\Im \Sigma^R(\omega)$ in the GK phase. 
The power-law exponents $s^\prime$ and $\kappa^\prime$ are obtained by fitting the AMEA results with \eqs{eq:PL_A} and \eqref{eq:PL_Sigma} on the interval $\Omega$. 
The same procedure applied to the HMPS results yields quite the same exponents (up to a deviation of $\approx 0.01$).
A power-law $\propto |\omega|^{-s}$ is plotted for comparison, see \eq{eq:PL_A}.
The error shadings, hardly to be seen in this figure, are estimates of the PH symmetry errors, see \app \ref{sec:error_estimation}.
}
\label{fig:eq_expos}
\end{figure}

The equilibrium case has been extensively studied in the literature.\cite{wi.fr.90,ch.ja.95,go.in.96,inge.96,bu.pr.97,go.in.98,lo.gl.00,gl.lo.00,bu.gl.00,in.si.02,gl.lo.03,vo.fr.04,fr.vo.04,fr.fl.06,le.bu.05,bu.co.08,gl.ki.11,ka.ma.12,fr.vo.13,aono.13}
It is well established that in a certain range of $r$, $U$ and $\Gamma$ the system displays a Kondo-like behavior, the so-called generalized Kondo effect.
In the GK phase, the spectral function and the retarded self energy are supposed to show a power-law behavior at small frequencies $|\omega|$,\cite{gl.lo.00}
\begin{align}
 A(\omega) &\propto |\omega|^{-s}, \phantom{1^\kappa} s=r \,, \label{eq:PL_A}\\
 \Im \Sigma^R(\omega) &\propto |\omega|^{\kappa}, \phantom{1^{-s}} \kappa > r \,. \label{eq:PL_Sigma}
\end{align}
First, we would like to address the question, how these properties are affected by the fact that AMEA cannot reproduce the pseudogap exactly down to asymptotically low energies.
Therefore, we study one set of parameters in the GK phase, according to the phase diagram in Fig.~5, \tcite{bu.gl.00}, which is reproduced in \fig{fig:phase_transition} of the present paper.
Specifically, we solve the many-body problem for $r=0.25$, $U=6$ and $\Gamma = 1$ (and a small temperature $T=0.05$) and compute the spectral function and retarded self energy.
Then we fit \eqs{eq:PL_A} and \eqref{eq:PL_Sigma} to these quantities on the interval $\Omega$ and extract the corresponding power-law exponents.
In the following, we denote their numerical values as $s^\prime$ and $\kappa^\prime$, respectively. 
The results are plotted in \fig{fig:eq_expos} together with the ones obtained by an HMPS treatment of the model for $T=0$ and $\eta = 0.1$.

From the results plotted in \fig{fig:eq_expos} we conclude that exponents extracted from the two methods, 
AMEA and HMPS, agree quite well.
We can also see that $\kappa^\prime > r$ is fulfilled, but $s^\prime$ is significantly larger than the predicted value $r$. 
This is, because the interval $\Omega$ used to determine the exponent lies at too large frequencies $|\omega|$.\footnote{
This is checked easily by calculating the $U=0$ spectral function for the exact physical hybridization function. 
The outcome shows that we need a good representation of the power-law exponent in $\Im \Delta^R$ down to $|\omega|$ values that are at least $1-2$ orders of magnitude smaller than the lower edge of $\Omega$ and this is not feasible within AMEA, neither HMPS at the moment.
}
On the other hand, it is not reasonable to go to smaller $|\omega|$ values, since the power-law is not well represented there in the hybridization function, see \fig{subfig:DeltaR_eq}.
Possibly, a more appropriate way to proceed here would be to use a logarithmic energy discretization as in NRG. 
However, without the possibility to integrate out high-energy degrees of freedom, this is no use here, and indeed the AMEA fit becomes quite unstable. 

It is also well established in the literature that upon increasing $U$, the system undergoes a QPT from the GK to an LM phase, where the Kondo-like behavior is absent.
Our next goal is to reproduce the phase boundary from Fig.~5 in \tcite{bu.gl.00}, i.e.~to numerically calculate the critical value $U_c$, which depends on $r$ and $\Gamma$, see \fig{fig:phase_transition}.
We would like to exploit \eqs{eq:PL_A} and \eqref{eq:PL_Sigma} for that purpose.
Since we find that it is difficult to extract the correct exponent $s'$ from the impurity spectral function, we choose to use the one of the self energy $\kappa '$, instead.
In \tcite{gl.lo.00} it is shown that, in the GK phase, this exponent must be larger than $r$.
In the equilibrium case, it is convenient to use the HMPS solver rather than AMEA for the numerical calculations.
In \figs{fig:eq_fit} and \ref{fig:eq_expos} we have already checked that both methods provide essentially the same values for the exponents up to a small deviation ($\approx 0.01$).
The HMPS solver is suitable for the equilibrium case, and, since it is based on a Hamiltonian time evolution, it is easier to employ and a bit faster, even for this large number of $1301$ bath sites.  

\begin{figure}[h]
\subfigure[]{
  \label{subfig:extrapol_eta_to_zero}
  \begin{minipage}[b]{\pw\columnwidth}
    \centering \includegraphics[width=1\textwidth]{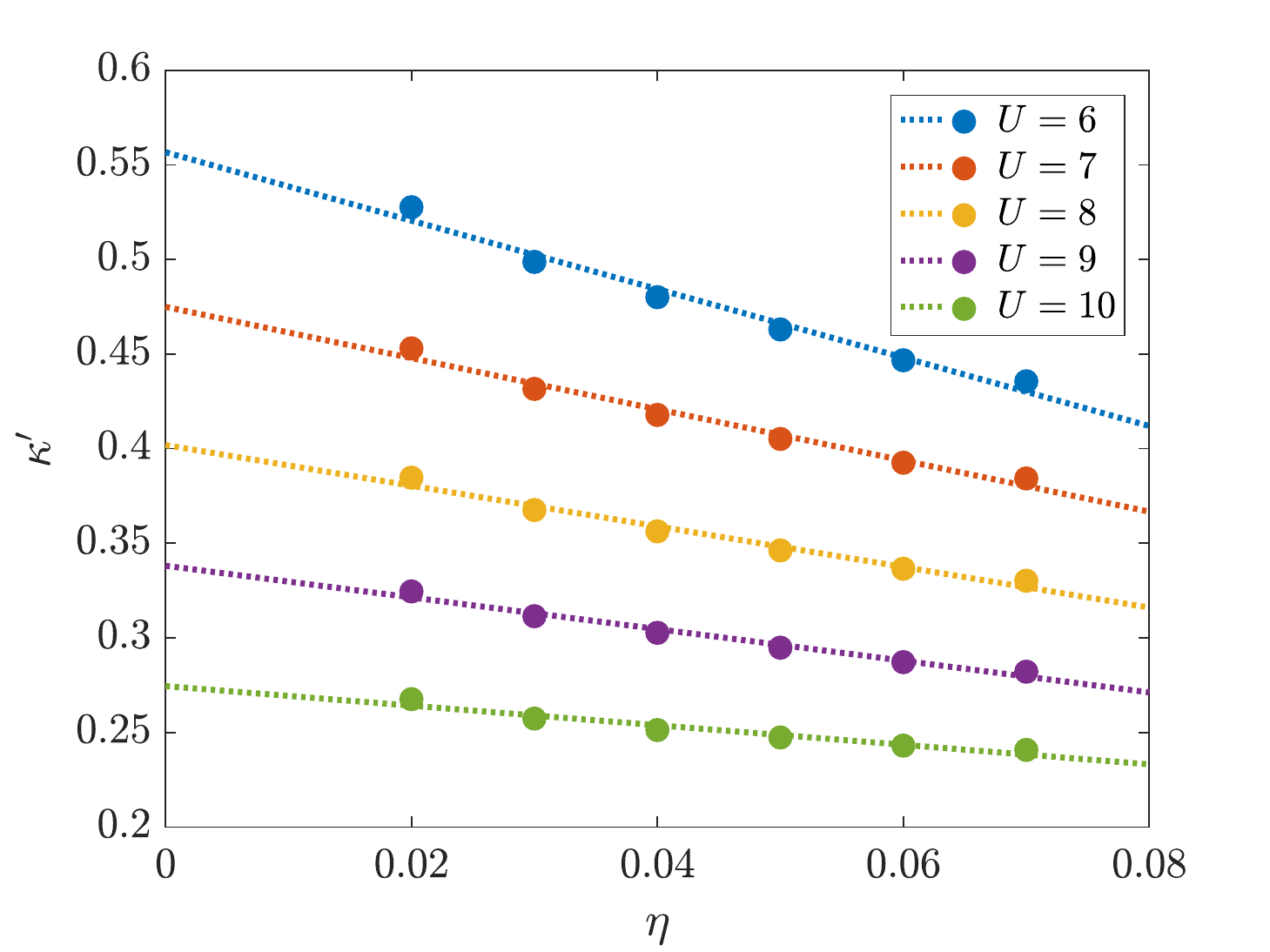}
  \end{minipage}
}\\
\subfigure[]{
  \label{subfig:extrapol_kappa_to_r}
  \begin{minipage}[b]{\pw\columnwidth}
    \centering \includegraphics[width=1\textwidth]{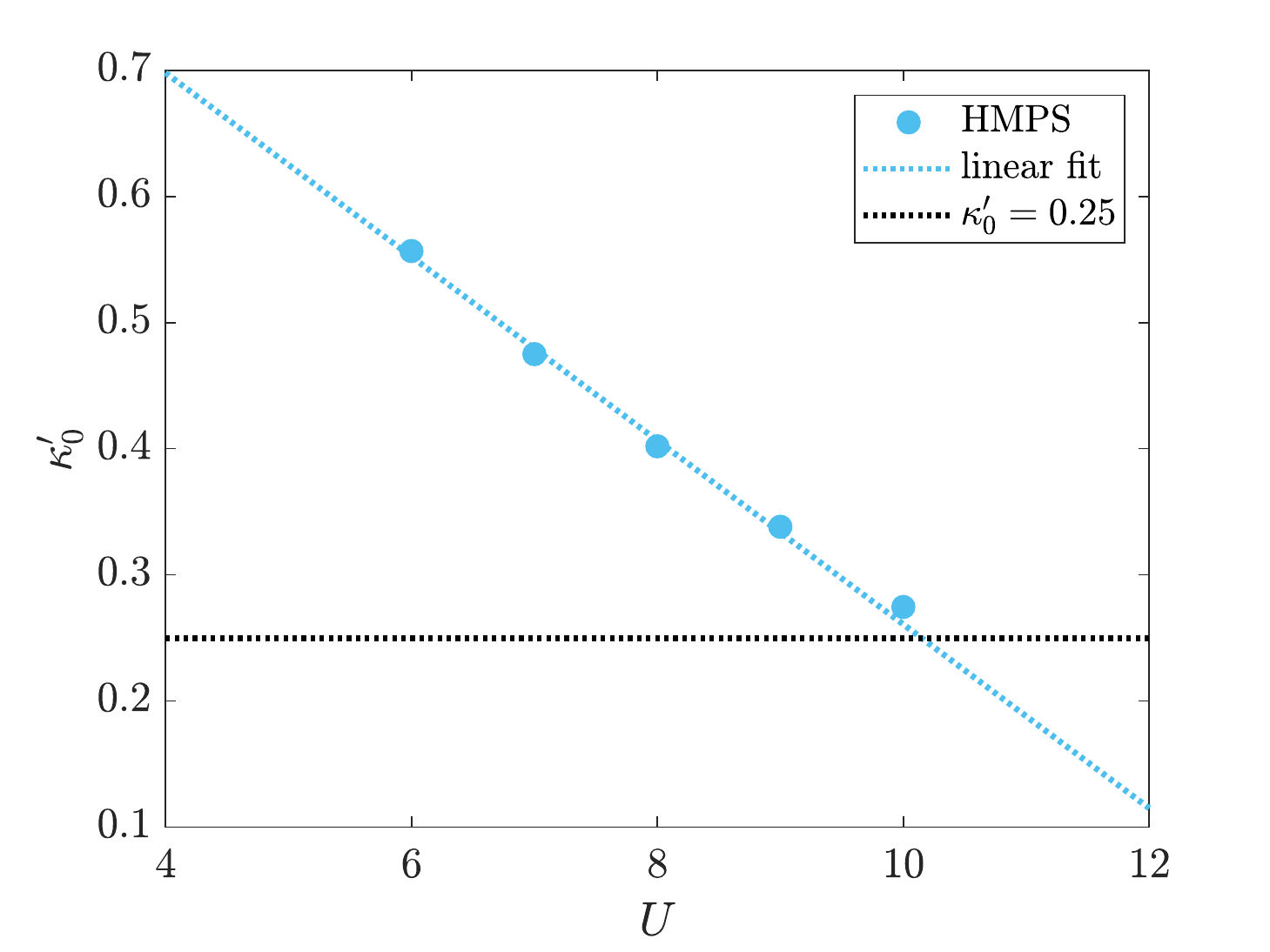}
  \end{minipage}
}\\
\caption{
Determination of the phase boundary by linear extrapolation of the power-law exponent $\kappa^\prime$ of the HMPS self energy in the GK phase.
First, \subref{subfig:extrapol_eta_to_zero} $\kappa^\prime$  is extrapolated to 
vanishing values of the broadening $\eta$ to extract
$\kappa^\prime_0 = \kappa^\prime(\eta \to 0)$ for various values of the interaction strength $U$. 
Second, \subref{subfig:extrapol_kappa_to_r} the critical interaction strength is determined from a second extrapolation, $U_c = U(\kappa^\prime_0 \to r)$. 
}
\label{fig:eq_extras}
\end{figure}

Specifically, we compute the Green's functions for different values of the interaction strength $U$ and extract the corresponding self energy from Dyson's equation \eqref{eq:total_Dyson} for various Lorentzian broadenings $\eta$. 
Then we fit $\Im \Sigma^R(\omega)$ on $\Omega$ and determine $\kappa^\prime$ as a function of $\eta$.
The results of this procedure are illustrated for $r=0.25$ in \fig{subfig:extrapol_eta_to_zero}.
We can see that $\kappa^\prime$ displays a significant dependence on $\eta$ (in contrast to $r^\prime$ and $s^\prime$)\footnote{
This is, due to the fact that the self energy is extracted from an inversion of the Green's function.
}
and that it is almost a linear function of $\eta$ for all considered values of $U$.
In order to extract the result without artificial broadening, we perform a linear extrapolation, $\kappa^\prime_0 = \kappa^\prime(\eta \to 0)$.
In \fig{subfig:extrapol_kappa_to_r} the obtained values for $\kappa^\prime_0$ are plotted and we find again an almost linear dependence on the interaction strength. 
According to the condition in \eq{eq:PL_Sigma}, the system should leave the GK phase at the value of $U$ for which $\kappa_0^\prime = r$. 
Thus, we perform a second linear extrapolation to extract the critical interaction strength as $U_c = U(\kappa_0^\prime \to r)$.

The phase boundary estimated in this way agrees well with the ones obtained by the numerical renormalization group, see \fig{fig:phase_transition}. %
In particular, the deviations within the results obtained from different NRG calculations are of the same size as the deviation of the HMPS results from the NRG results for the considered values of $r$.\footnote{
Even though the phase diagram of \tcite{bu.gl.00} was obtained under the assumption $U \ll D$, where $D$ is the bandwidth, while in this work we have $U \lesssim D$.
For $U \ll D$, $D$ is irrelevant as energy scale and the phase boundary is solely determined by $\Gamma$, $U$ and $r$, see \tcite{bu.gl.00}.
}
It is notable, though, that the HMPS scheme tends to overestimate the critical interaction strength, yielding slightly smaller values $U_c^{r-1}$ in \fig{fig:phase_transition}.
This could be improved by taking into account that $\kappa^\prime(\eta)$ is not strictly a linear function. 
By accounting for its curvature, one obtains slightly larger values $\kappa_0^\prime$ (see \fig{subfig:extrapol_eta_to_zero}).
This, in turn, results in smaller critical interaction strengths (see \fig{subfig:extrapol_kappa_to_r}) and thus in larger values of $U_c^{r-1}$, closer to the corresponding NRG results.
From the literature it is known that the GK phase can occur only for $0<r<0.5$, see e.g.~\tcite{fr.vo.04}. 
Close to the phase boundary at $r \to 0.5$, the HMPS calculations are more involved, the quantities $\kappa^\prime(\eta)$ and $U(\kappa_0^\prime)$ are much more difficult to obtain and the extrapolation scheme described above breaks down.
Therefore, in \fig{fig:phase_transition} the HMPS results are plotted only up to $r = 0.45$.

\begin{figure}[h]
  \begin{minipage}[b]{\pw\columnwidth} %
    \centering \includegraphics[width=\textwidth]{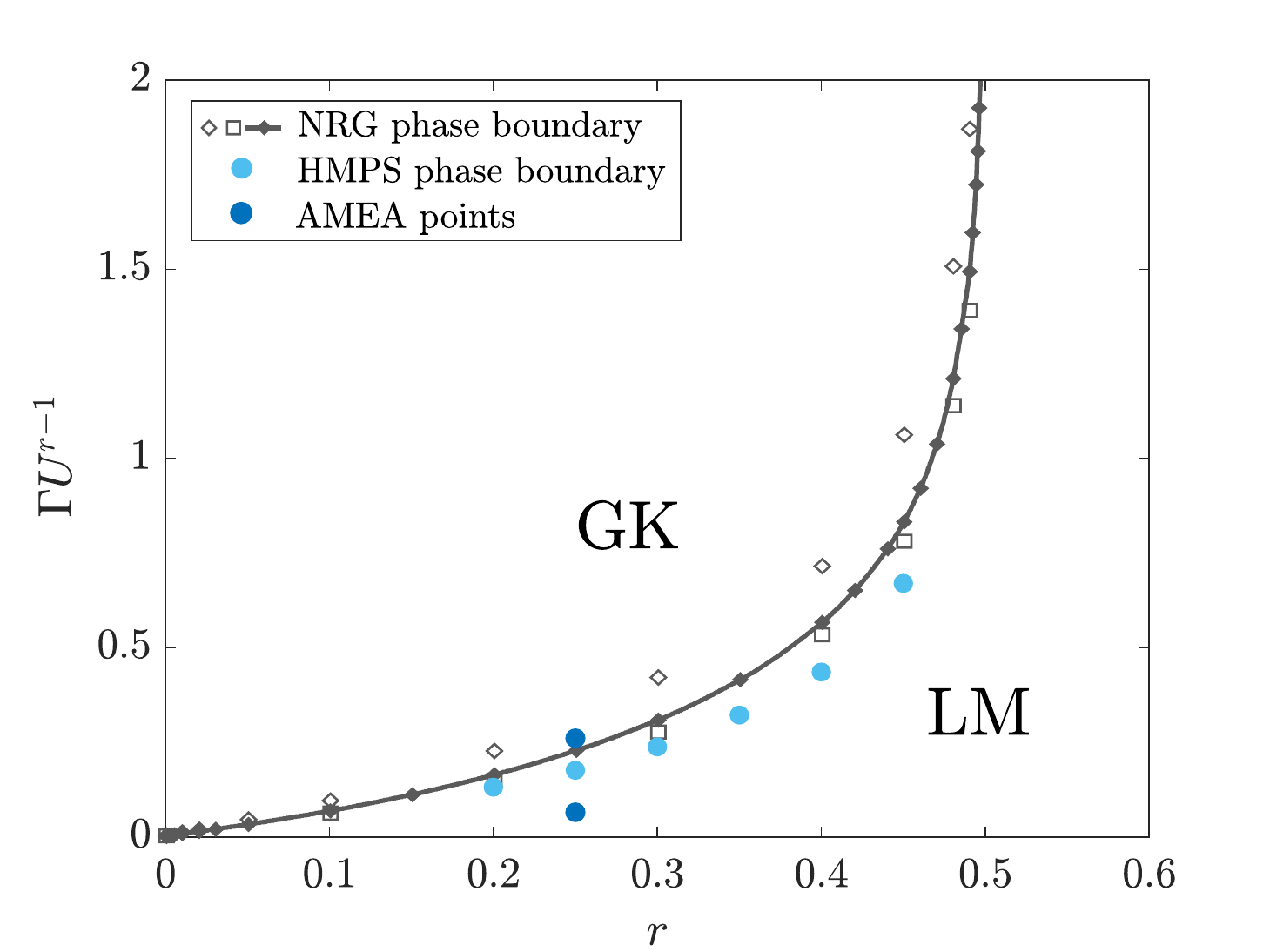} 
  \end{minipage}
\caption{
Phase diagram adapted from \tcite{bu.gl.00} (with kind permission) displaying different NRG results.\protect\footnote{Results obtained by the local moment approach were removed here, since they are not relevant to the present discussion.}
On top of this we present our HMPS results for the phase boundary obtained via the extrapolation scheme discussed in the text.
We also indicate the two points we consider in AMEA, i.e. $r=0.25$, $U=6$ and $\Gamma=0.25$ and $\Gamma=1$. 
If $U$ is much smaller than the bandwidth, the phase boundary for a given $r$ is expected to depend on  $\Gamma U^{r-1}$ only. \cite{bu.gl.00}
}
\label{fig:phase_transition}
\end{figure}

It is remarkable that our results reproduce  the NRG phase boundary to this level of accuracy, even though the low energy part of the bath hybridization function used in our calculation is not reproduced perfectly and the Kondo effect is of course especially dependent on the hybridization function at $\omega \approx 0$.
The encouraging performance of the HMPS scheme and the good agreement between the results obtained from HMPS and from AMEA prompts us to use AMEA to study the system in its nonequilibrium steady state, for which HMPS cannot be used.
\subsubsection{Nonequilibrium steady state}
\label{subsubsec:nonequilibrium}
\begin{figure*}[!t]
\subfigure[]{
  \label{subfig:A_neq_Kondo}
  \centering \includegraphics[width=0.82\textwidth]{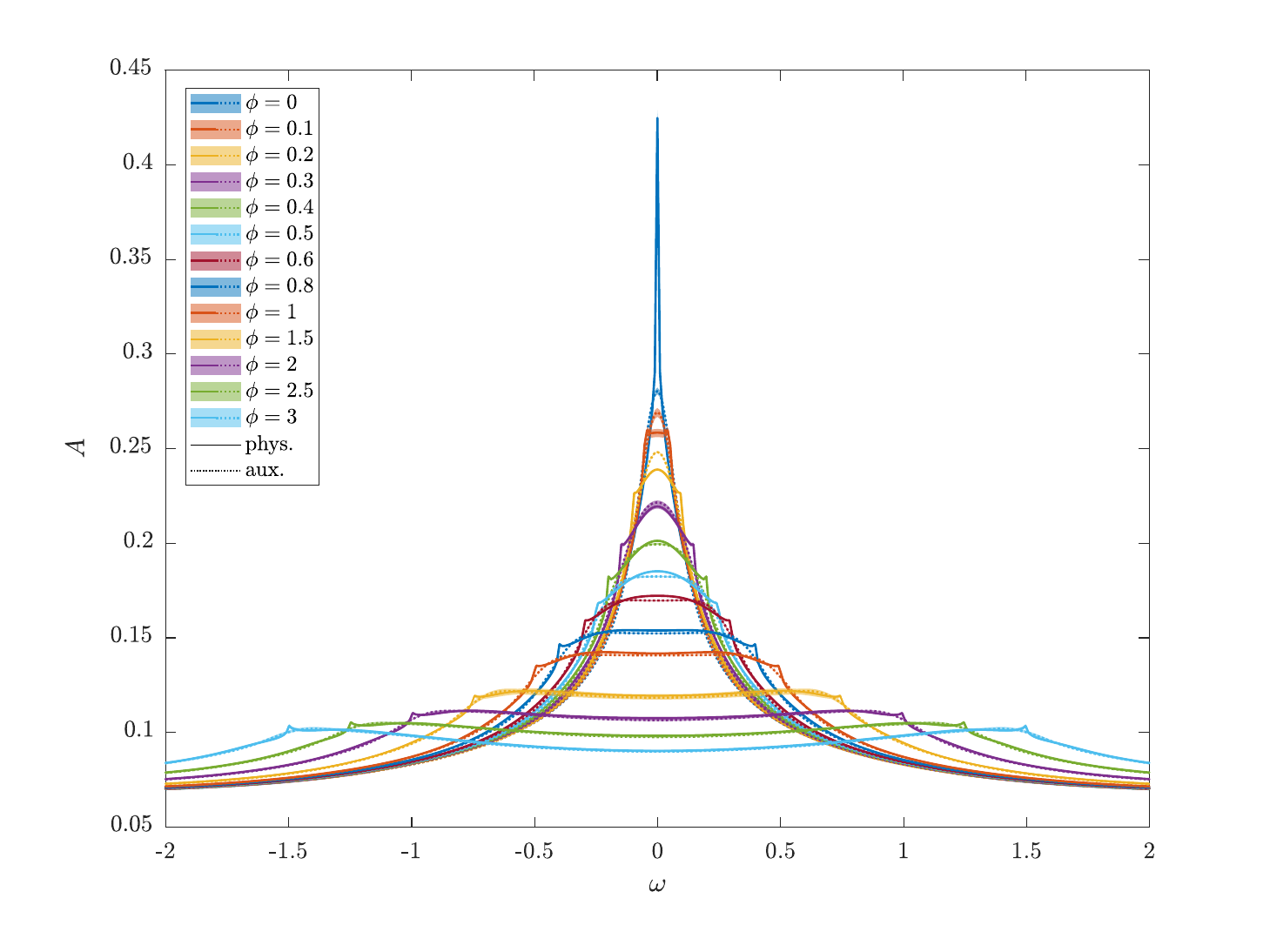}  %
}
\subfigure[]{
  \label{subfig:Sigma_neq_Kondo}
  \centering \includegraphics[width=0.45\textwidth]{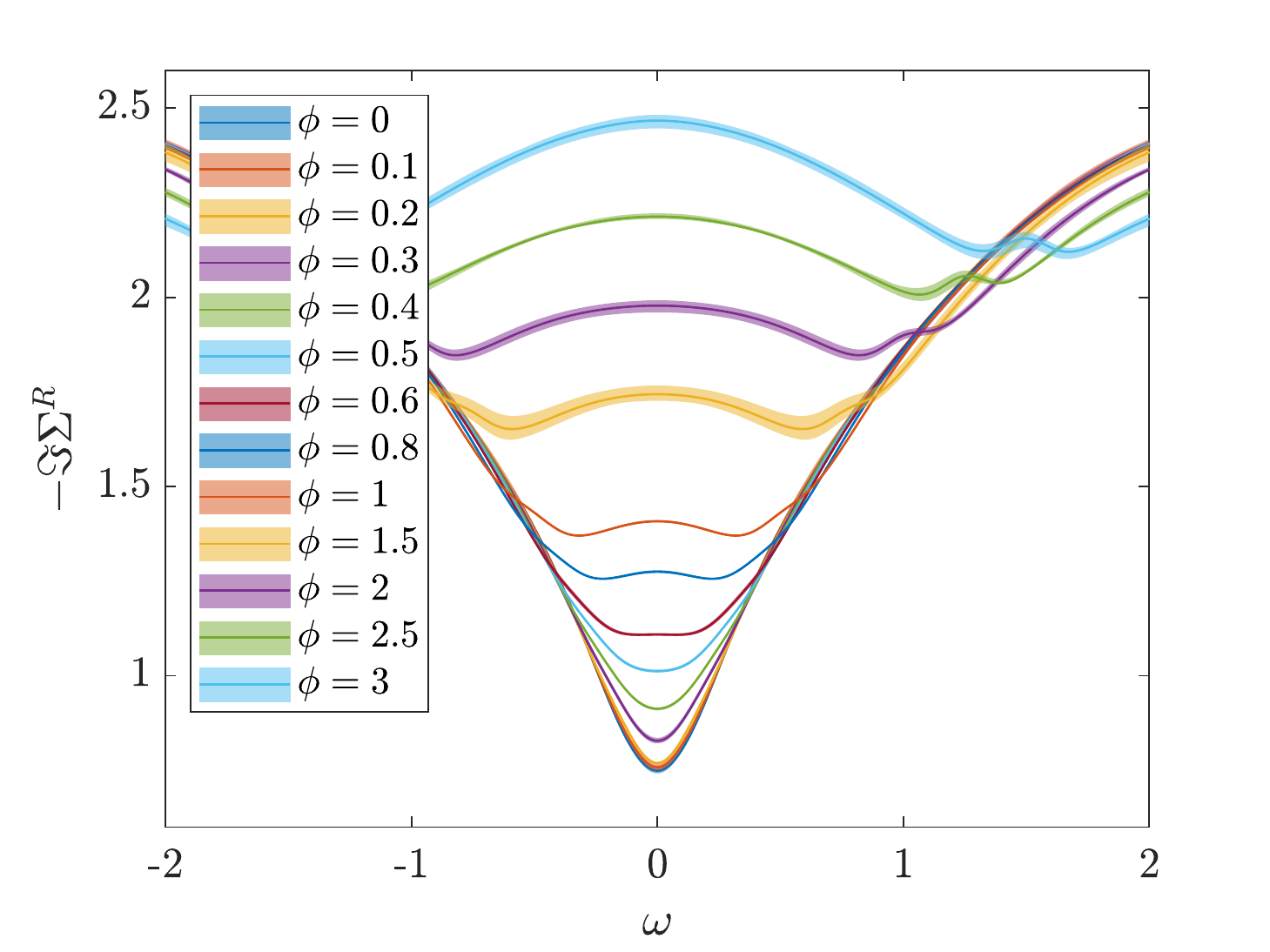} %
}
\subfigure[]{
  \label{subfig:G_neq_Kondo}
  \centering \includegraphics[width=0.45\textwidth]{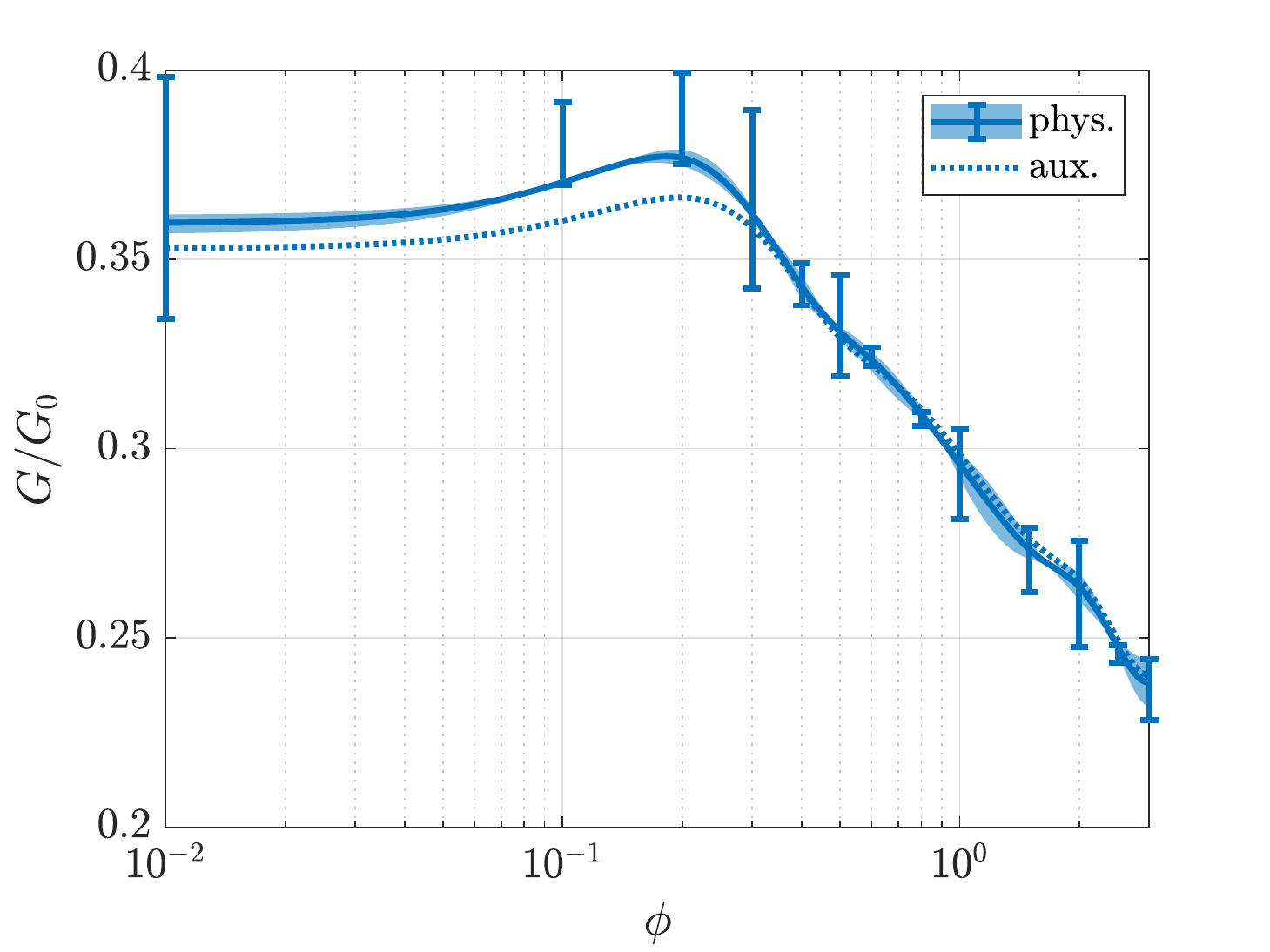} %
}
\caption{
Nonequilibrium ($\phi > 0$) quantities in the Kondo regime, \subref{subfig:A_neq_Kondo} spectral function, \subref{subfig:Sigma_neq_Kondo} retarded self energy, \subref{subfig:G_neq_Kondo} differential conductance.
The solid lines are the physical quantities and the dotted lines the auxiliary ones, see \se \ref{subsubsec:phys_vs_aux}.
Notice that the two curves are often indistinguishable.
The error shadings and error bars are estimates based on symmetry considerations, see \app \ref{sec:error_estimation}.
}
\label{fig:neq_Kondo}
\end{figure*}
\begin{figure*}[!t]
\subfigure[]{
  \label{subfig:A_neq_LM}
  \centering \includegraphics[width=0.82\textwidth]{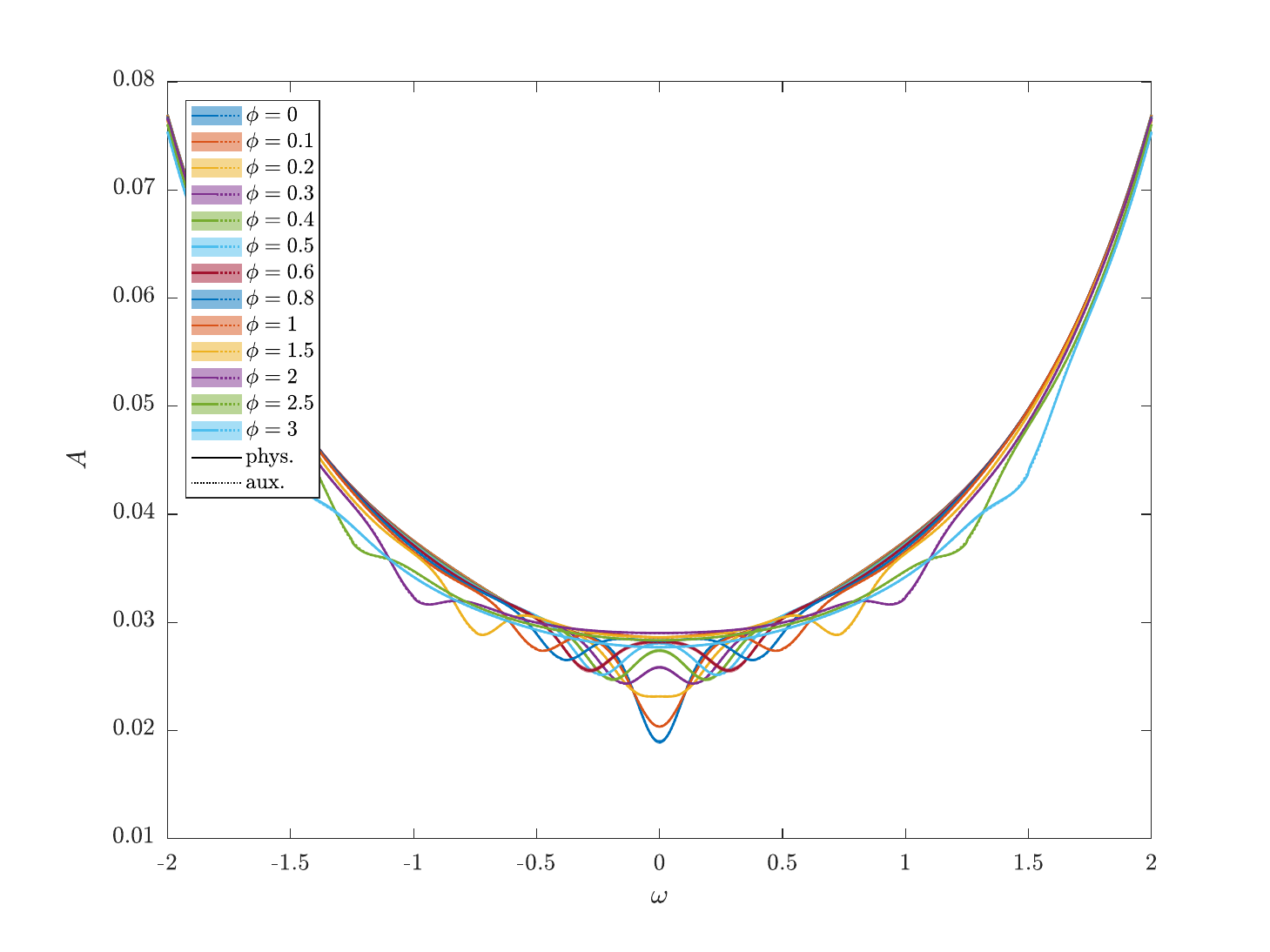}  
}\\
\subfigure[]{
  \label{subfig:Sigma_neq_LM}
  \centering \includegraphics[width=0.45\textwidth]{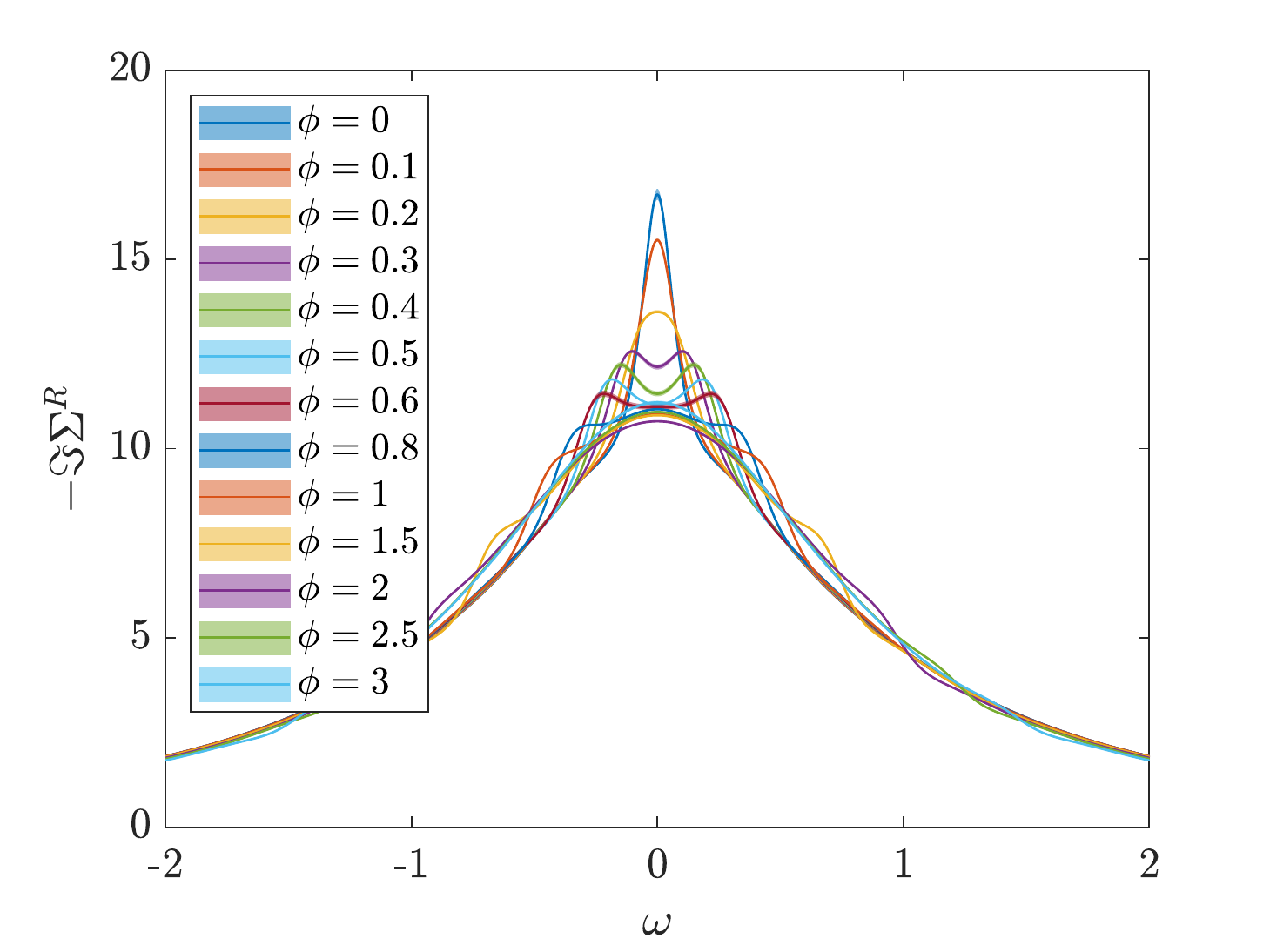} 
}
\subfigure[]{
  \label{subfig:G_neq_LM}
  \centering \includegraphics[width=0.45\textwidth]{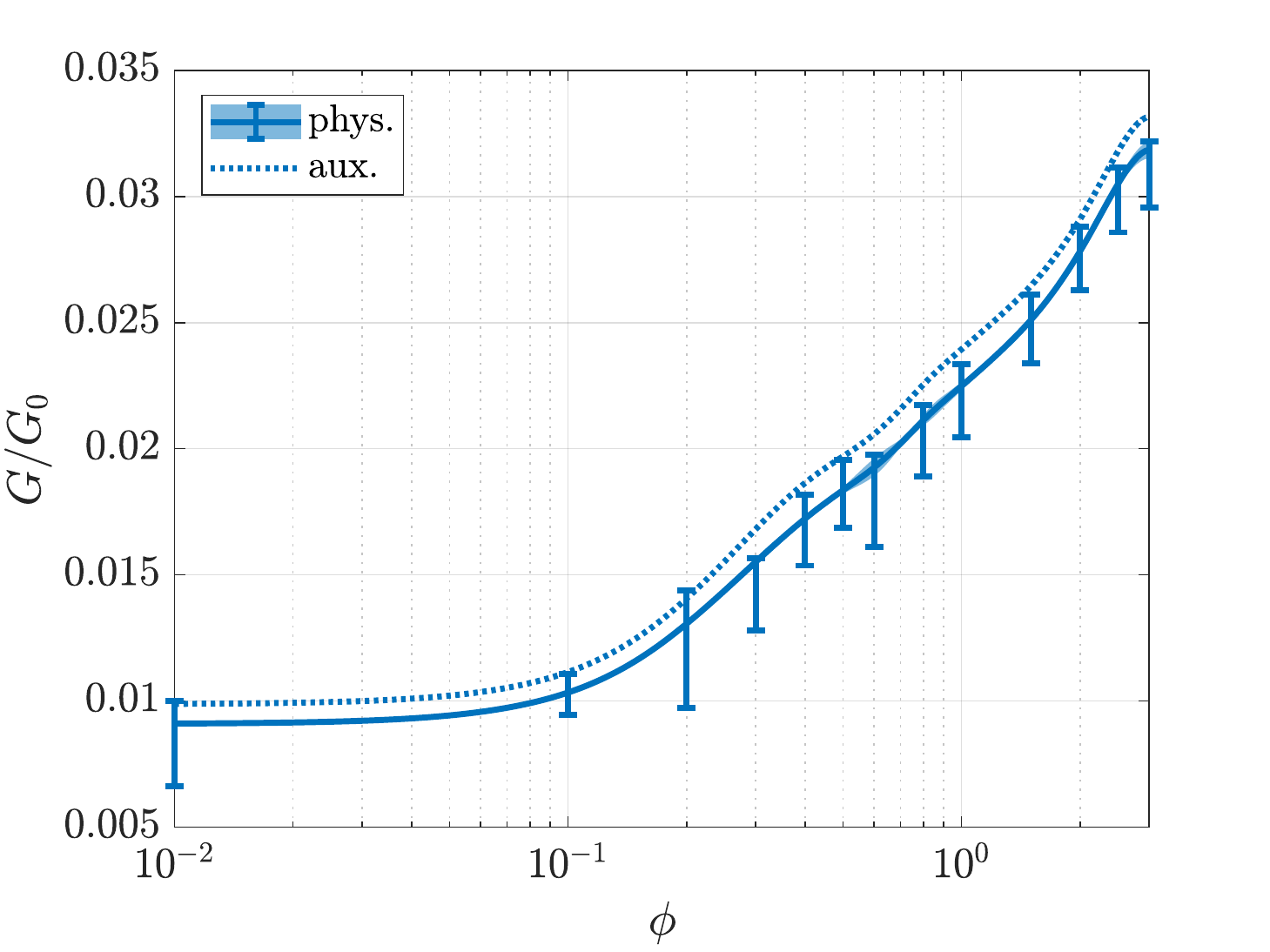} 
}
\caption{
Nonequilibrium ($\phi > 0$) quantities in the LM regime. Conventions are as in \fig{fig:neq_Kondo}.
}
\label{fig:neq_LM}
\end{figure*}

We now present nonequilibrium steady state results obtained by applying a finite bias voltage.
Since the calculations are more demanding than the conventional HMPS ones, we focus on two points in the (equilibrium) phase diagram \fig{fig:phase_transition}, one in the GK and another in the LM phase, instead of doing a complete sweep of parameters. 
Specifically, we take $r=0.25$, $T=0.05$, $U=6$, and $\Gamma=0.25$ and $1$, respectively. 

We start by studying the behavior of the Kondo peak as a function of voltage. %
Therefore, we plot in \fig{subfig:A_neq_Kondo} and \subref{subfig:Sigma_neq_Kondo} the spectral function and the imaginary part of the self energy.
In the Kondo regime, we observe that upon increasing the bias voltage from $\phi=0$ the equilibrium Kondo peak is suppressed and broadened and, at some value of the voltage, it splits in two peaks.
The split peaks then move apart together with the chemical potentials and they are further suppressed and broadened.
Qualitatively, this is very similar to the situation observed for the nonequilibrium SIAM without a pseudogap.\cite{do.nu.14,do.ga.15,ande.08,wi.me.94,le.sc.01,ro.kr.01,fu.ue.03,nu.he.12,co.gu.14,do.ga.16,ha.he.07}
In our data, the splitting becomes visible for $\phi \geq 0.8$ in the spectral function and, even before, for $\phi \geq 0.6$ in the self energy.

A measure for the accuracy of the mapping between \eq{h} and the auxiliary open system, which is at the basis of the AMEA approach, can be read off from the deviations between the physical and the auxiliary spectral functions, defined in \se \ref{subsubsec:phys_vs_aux}. 
In the limit in which the mapping to the auxiliary system becomes exact, i.e. for large $N_B$, these quantities become identical.
Our data show that $A_\mathrm{aux}$ and $A_\mathrm{phys}$ differ only slightly for $\phi \geq 0.3$.  
Decreasing the voltage below $\phi = 0.3$ increases this deviation, especially for $\omega$  between the chemical potentials, and it is largest at $\phi = 0$, where the exact physical spectral function is expected to diverge at $\omega = 0$. 
Here we expect the accuracy of the AMEA mapping to be less reliable.

It is notable that as soon as the Kondo split peaks appear, they are very broad and poorly defined, even more in $A_\mathrm{phys}$, but also in $A_\mathrm{aux}$.
They are first located at $|\omega|$ values slightly below $|\mu_\lambda|=\phi/2$, which they reach monotonically upon increasing the bias voltage.
The physical spectral function displays additional features, namely two cusps at $\pm \phi/2$, not to be confused with the Kondo split peaks.
We believe these to be artefacts originating from the difference between the auxiliary and the physical system and 
we expect them to disappear upon improving the accuracy.

\figs{subfig:A_neq_LM} and \subref{subfig:Sigma_neq_LM} are obtained for the same parameters as \figs{subfig:A_neq_Kondo} and \subref{subfig:Sigma_neq_Kondo}, but a reduced hybridization strength of $\Gamma = 0.25$, instead of $\Gamma = 1$.
According to the phase diagram in \fig{fig:phase_transition}, the equilibrium system is in the LM phase, here. 
This is confirmed by our results which, indeed, do not show signatures of the Kondo effect anymore, neither in equilibrium nor at finite bias voltage.\footnote{
Notice that it is not clear, whether a true phase transition or rather a crossover occurs between the two phases at finite voltage.
}
Specifically, at nonzero $\phi$, we observe dips in the spectral function located almost exactly at the values of the chemical potentials that appear to emerge as images of the dips in the 
leads' density of states.
Also in this case, the physical and auxiliary spectral functions agree very well with each other, thus making us confident about the accuracy of our results.
Artificial cusps at $|\mu_\lambda|$ are also present in $A_\mathrm{phys}$, but they are much smaller than the cusps in the Kondo regime.\footnote{
Here, a smaller truncated weight was chosen in the SVDs in the MPS time evolution, which could explain this improved accuracy.
}
These essentially lie within the error shadings of $A_\mathrm{aux}$ and are notable only upon zooming in. 

\figs{subfig:G_neq_Kondo} and \ref{subfig:G_neq_LM} display the differential conductance $G$, obtained from \eqs{eq:curr_formula} and \eqref{eq:G_general_formula} as a function of the bias voltage at parameters corresponding to the Kondo and the LM regime.
A notable difference with respect to the conventional SIAM is that in the Kondo regime, the maximum of $G(\phi)$ appears to be shifted to a finite voltage of $\phi \approx 0.2$.
On the other hand, for $\phi \gtrsim 0.2$, $G(\phi)$ decreases logarithmically, as usual. 
The unusual structure of the differential conductance in the Kondo regime is probably, due to the fact that the position of the pseudogap is shifted along with the bias voltage. 
On the other hand, one should be aware of the fact that, due to the relatively large error bars,\footnote{
The error bars as well as the error shadings are estimated from the violation of PH symmetry of the corresponding quantities, as discussed in \app \ref{sec:error_estimation}.
Violation of PH symmetry via protocol 2 produces a slight difference between the left- and right-moving current $|j_L|$ and $|j_R|$, which is clearly unphysical for the steady state.
Since $G(\phi)$ is obtained by numerical differentiation of the $j_\lambda(\phi)$ curves, its error is amplified. 
This explains, why the error bars in the $G(\phi)$  are so large.
}
it is not clear, whether the maximum at finite voltage is a genuine feature: 
strictly speaking, also a maximum at $\phi=0$ would be consistent with the error bars.
Furthermore, we already noticed in \fig{subfig:A_neq_Kondo} that the deviations between $A_\mathrm{phys}$ and $A_\mathrm{aux}$ are large at $\phi \leq 0.2$ compared to the other bias voltages and this is exactly, where the peculiar behavior of $G(\phi)$ sets in.
In contrast  to the Kondo regime, \fig{subfig:G_neq_LM} shows that in the LM regime the differential conductance increases with the bias voltage, as expected.

\begin{figure}[h]
  \begin{minipage}[b]{\pw\columnwidth} %
    \centering \includegraphics[width=1\textwidth]{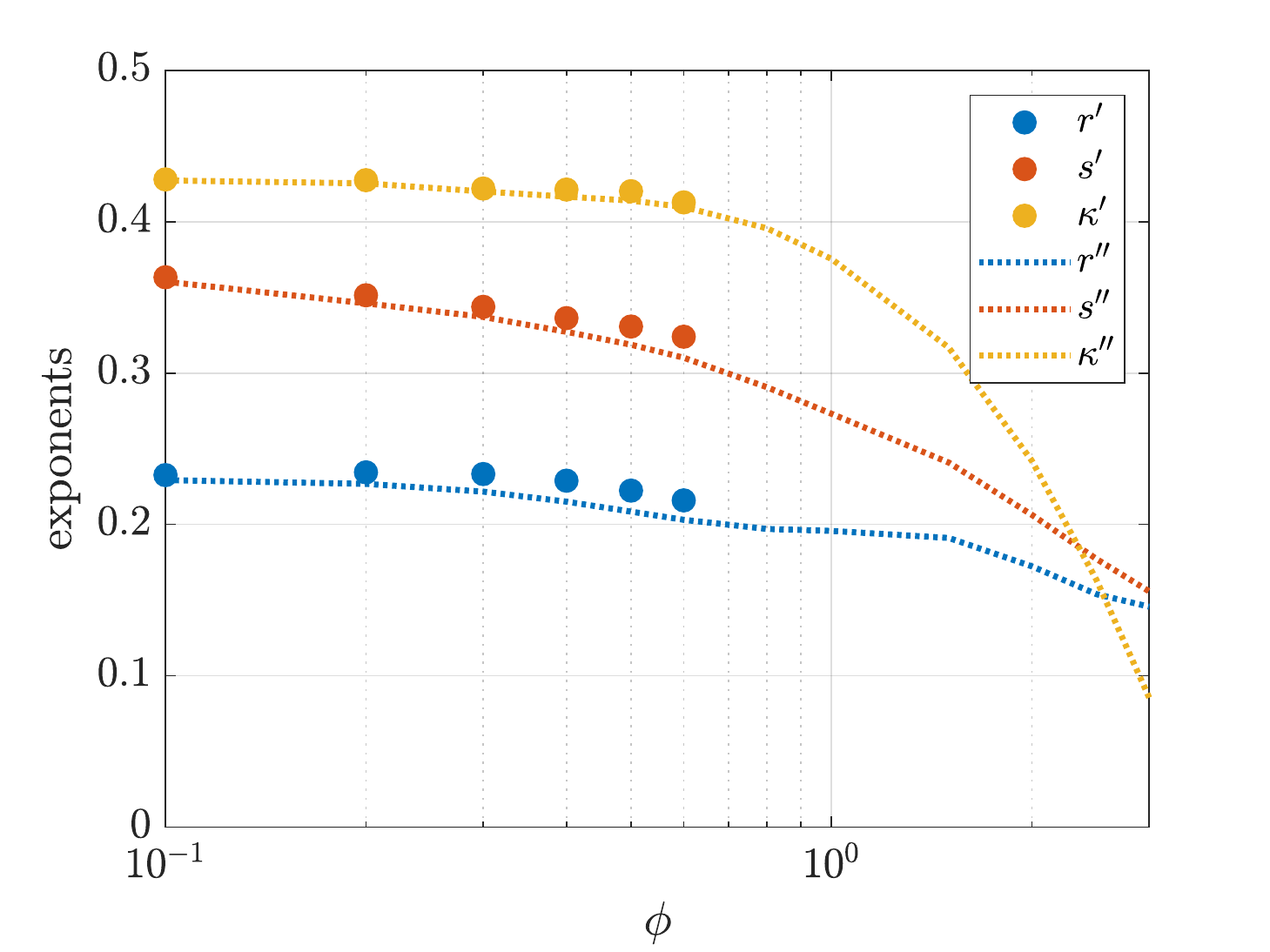}
  \end{minipage}
\caption{
Nonequilibrium ($\phi > 0$) effective power-law exponents as a function of the bias voltage $\phi$.
The three pairs of exponents are extracted from a fit of the auxiliary retarded hybridization function ($r^\prime,r^{\prime\prime}$), the spectral function ($s^\prime,s^{\prime\prime}$) and the retarded self energy ($\kappa^\prime,\kappa^{\prime\prime}$) with \eq{eq:PL_Delta_neq}-\eqref{eq:PL_Sigma_neq}. 
The single and double primes correspond to different fitting intervals $\Omega(\phi)$ and $\Omega_1(\phi)$, see text.
}
\label{fig:neq_expos}
\end{figure}

We now attempt at extracting ``effective'' power-law exponents in the Kondo regime, as we do in equilibrium, by carrying out a fit of the nonequilibrium curves.  
More specifically, in analogy to \eqs{eq:PL_Delta}, \eqref{eq:PL_A} and \eqref{eq:PL_Sigma}, we fit the behavior
\begin{align}
 \Im \Delta^R(\omega) &\propto |\omega-\mu_L|^{r\hphantom{-}} + |\omega-\mu_R|^{r} \,, \label{eq:PL_Delta_neq}\\
 A(\omega) &\propto |\omega-\mu_L|^{-s} + |\omega-\mu_R|^{-s} \,, \label{eq:PL_A_neq}\\
 \Im \Sigma^R(\omega) &\propto |\omega-\mu_L|^{\kappa\hphantom{\!-}} + |\omega-\mu_R|^{\kappa} \,. \label{eq:PL_Sigma_neq}
\end{align}
The finite voltage and the imperfect pseudogap set a low-frequency cutoff to this behavior, which we expect not to hold down to zero frequency.
The exponents, $r^\prime(\phi)$, $s^\prime(\phi)$ and $\kappa^\prime(\phi)$, obtained by a fit on the interval $\Omega(\phi)$, defined in \eq{eq:Omega_phi_int}, are presented in \fig{fig:neq_expos}.
Since this interval shrinks upon increasing the bias voltage, we can faithfully perform the fit only for voltages $\phi \lesssim 0.6$, as discussed below \eq{eq:Omega_phi_int}.
Thus, we can just catch the beginning of the interesting voltage region, where the Kondo split peaks start developing at $\phi \approx 0.6$, according to \fig{subfig:A_neq_Kondo}. 
Moreover, due to the lower cutoff in energy, the extracted exponents can only provide a rough semi-quantitative estimate.
In the range $\phi \lesssim 0.6$ the exponents depend only slightly on the bias voltage. 
Nevertheless, it is notable that $r^\prime(\phi)$ and $s^\prime(\phi)$ are almost parallel.
This may indicate that deviations in $\Im \Delta^R(\omega)$ (such as between $\Im \Delta^R_{\mathrm{aux}}$ and $\Im \Delta^R_{\mathrm{phys}}$) mainly translate into deviations in the spectral function, affecting $\Im \Sigma^R(\omega)$ in a minor way.\footnote{
This argument is supported by the fact that \underline{$\Delta$}\,\! enters \underline{$G$}$_0$\,\! and \underline{$G$}\,\! in the same way in Dyson's equation.
This is easily seen by comparing the general form of \eq{eq:total_Dyson} to the result for zero self energy, \underline{$G$}$\,=\,$\underline{$G$}$_0$.
}
Indeed, if the self energy is more stable against numerical inaccuracies than the spectral function, one could try to exploit this to study the phase transition or crossover also out of equilibrium, with a scheme similar to the one presented in \se \ref{subsubsec:equilibrium}.
However, in order to do this, it would be necessary to resolve a larger fraction of the interesting voltage region, $\phi \gtrsim 0.6$, which, on the other hand, would require a larger $\Omega$ interval, where the power-law in the auxiliary hybridization function is accurately resolved. 

\fig{fig:neq_expos} also displays the power-law exponents $r^{\prime\prime}(\phi)$, $s^{\prime\prime}(\phi)$ and $\kappa^{\prime\prime}(\phi)$ fitted on a larger interval $\Omega_1(\phi) = 0.2 + \frac{\phi}{2} < |\omega| < 1.2 + \frac{\phi}{2}$, which is obtained by a rigid shift of the equilibrium interval $\Omega$ by $\frac{\phi}{2}$. 
In the region $\phi\lesssim 0.6$, where both kinds of exponents ($\prime$ and $\prime\prime$) are defined, their values lie very close to each other.
This confirms that the influence of the exponential factor in the hybridization function is negligible on these frequency and voltage intervals. 
\section{Summary and Conclusions}
\label{sec:conclusion}

In this work we addressed the single-impurity Anderson model with leads displaying a power-law pseudogap in the density of states (PSG SIAM) by means of a nonperturbative approach to deal with nonequilibrium steady states, the auxiliary master equation approach (AMEA).
We studied the generalized Kondo (GK) and the local moment (LM) phase of this model in equilibrium as well as their extension out of equilibrium.

In order to assess the validity of our approach, we first compared the results with the ones obtained with a direct MPS time evolution of the Hamiltonian (HMPS).\cite{ba.zi.17} 
HMPS is faster than AMEA and it can treat a larger number of bath sites in equilibrium, but, on the other hand, it cannot deal with a nonequilibrium steady state, due to the lack of a dissipation mechanism.
We found that the spectral function, the self energy and the power-law exponents of these quantities agree very well between AMEA and HMPS, see \fig{fig:eq_expos}.
Furthermore, we implemented a scheme to find the phase boundary upon linear extrapolation of the power-law exponent of the self energy in the GK phase.
The phase boundary obtained in this way agrees quite well with previous NRG results, see \fig{fig:phase_transition}.

Out of equilibrium, we observe a splitting of the Kondo peak in the spectral function and in the self energy as a function of the bias voltage, see \figs{subfig:A_neq_Kondo} and \subref{subfig:Sigma_neq_Kondo}, as in the case of the conventional Kondo effect.
On the other hand, the differential conductance appears to display a peculiar maximum at finite bias voltage, \fig{subfig:G_neq_Kondo}, which could be caused by the shift of the hybridization functions at finite bias voltages.
Due to the error bars, it is not clear, whether this maximum can be considered a genuine feature of the model.
We are not aware of any other work on this model displaying this feature. 
For example, in \tcite{ha.ko.15}, the conductance maximum occurs at zero bias. 
However, this work also does not show a splitting of the Kondo resonance at finite bias voltages. 
The authors 
attributed this to the fact that the system is not in the limit of large interactions.
A comparison with our results is difficult, since the position of the pseudogap as a function of voltage is considered differently in our paper. 
More specifically, while in \tcite{ha.ko.15} the pseudogap is fixed at $\omega=0$, in our case it moves with the chemical potentials of the two leads, consistent with a rigid shift of the two leads.

Strictly speaking, what we observe in the Kondo regime, is the result of a superposition of the (pseudogap) GK effect with a small contribution from the ordinary one. 
This is, due to the fact that the imperfect mapping produces a nonzero residual $\Gamma_\mathrm{resid} = -\Im \Delta^R_\mathrm{aux}(0) \approx 0.39 $, even at zero bias voltage.
However, the contribution from this residual DOS is negligible, since the resulting Kondo temperature $T_{K,\mathrm{resid}} \approx 0.0025$ is much smaller than the temperature of our data $T \approx 20 \, T_{K, \mathrm{resid}}$.\footnote{
$T_K$ is estimated  with the widely used formula from \tcite{hews}, $T_K = \sqrt{\Gamma U / 2} \, \exp{ \left[ - \pi U / (8 \Gamma) \right] }$, assuming a constant lead DOS with a hybridization strength of $\Gamma = \Gamma_\mathrm{resid}$.
}
Therefore, the Kondo resonances shown in \fig{subfig:A_neq_Kondo} and \subref{subfig:Sigma_neq_Kondo} are clearly dominated by the pseudogap GK effect.

It would be clearly desirable to be able to extend an accurate mapping of the hybridization function down to smaller $|\omega|$ values.
This would further reduce the contribution of the ordinary Kondo effect and it would allow for a more accurate analysis of the low-frequency behavior.
In previous works, \tcites{do.so.17,do.ga.15}, we demonstrated that the accuracy of the mapping increases exponentially upon increasing the number of bath sites.
However, this is only true, if we find good enough minima of the cost function measuring the difference between
$\underline{\Delta}_{\mathrm{aux}} $ and $\underline{\Delta}_{\mathrm{phys}} $. 
This has, so far, turned out to be difficult for the PSG model studied here.
In order to resolve the power-law with the cusp, bath sites on all energy scales would be required, as used in NRG. 
To make progress in this direction, we tried to fit the hybridization function on a logarithmic frequency grid and/or include its power-law exponent explicitly into the cost function, but without success so far.  The fit seems to be quite unstable in all of these cases. 

On the technical side, this work presents a development of the AMEA Lindblad many-body impurity problem within a matrix product states algorithm. 
Due to the reduced local Hilbert space obtained by separating the degrees of freedom, the present implementation is faster and more stable than the one of our previous work, \tcite{do.ga.15}.  
On the other hand, the disadvantage of the structure used here is that additional long-range couplings between the impurity and the baths are introduced, as illustrated in \fig{fig:sketch_tensor_network}, and the entanglement must be carried across the sites in between, which causes the bond dimension to increase. 
An obvious way to avoid this is a ``fork'' structure, in particular, a ``double fork'', which has three bonds at the impurity, instead of two. 
This structure naturally takes into account the spin separation as well as the separation between \almostfull and \almostempty baths and, at the same time, only has nearest neighbor couplings.
The disadvantage of this scheme is that it cannot be represented by MPS, because of the third bond, and it thus
requires the implementation of a new tensor network, similar to the one described in \tcite{ba.zi.17}. 
Work along these lines is in progress.
\begin{acknowledgments}

We would like to thank Franz Scherr for providing a first implementation of the AMEA mapping using the python library tensorflow.
This work was supported by the Austrian Science Fund (FWF) within the project P26508 and the START program Y746, as well as NaWi Graz.
The numerical results presented here have been carried out on the D-Cluster Graz and on the VSC-3 HPC Cluster Vienna.

\end{acknowledgments}

\appendix
\section{Construction of a nonequilibrium system from equilibrium bath parameters}
\label{sec:NEQ_system_construction}
Here, we show two results concerning the representation of a noninteracting fermionic bath in terms of Lindblad open systems, focussing on a geometry that is suitable for a treatment with MPS.
As discussed in our previous work,\cite{do.ga.15} for the sake of an MPS treatment, it is convenient to connect the impurity to a bath which is \almostfull and one which is \almostempty.
Each of the two baths should have a one-dimensional chain geometry and couple on each side of the impurity.
This geometry guarantees a slower propagation of entanglement.  
For this reason, in \app \ref{subsec:F_E_bath} we show, how to represent an arbitrary hybridization function as originating from a \almostfull and an \almostempty bath.
This is valid both for a nonequilibrium as well as for an equilibrium $\phi=0$ hybridization function. 
In our paper, it is convenient to start with such a representation for the fit of an equilibrium bath and then use this solution to produce a \almostfull-\almostempty representation for a finite voltage $\phi \not=0$. 
How this is done, is shown in \app \ref{subsec:EQ_bath_to_NEQ_bath}. 
\subsection{Splitting into a \almostfull and \almostempty bath}
\label{subsec:F_E_bath}
The effects of an arbitrary noninteracting fermionic bath on a single-site impurity are completely described by its hybridization function $\underline{\Delta}(\omega)$ in Keldysh space.
Here, we show that any (equilibrium or nonequilibrium) $\underline{\Delta}$ can always be split as $\underline{\Delta} = \underline{\Delta}_F + \underline{\Delta}_E$, where $\underline{\Delta}_F$ describes a \almostfull (F) and $\underline{\Delta}_E$ an \almostempty (E) (equilibrium) bath. 
As discussed above, these two baths are represented by a Lindblad equation, where $\boldsymbol{\Gamma}^{(1)}=0$ or $\boldsymbol{\Gamma}^{(2)}=0$, respectively. 

For better readability, we omit the frequency argument $\omega$ and introduce the two components of the hybridization function
\begin{equation}
\label{dri}
\Delta^{Ri} \equiv \Im \Delta^R \,, \quad\quad\quad \Delta^{Ki} \equiv \frac{\Delta^K}{2i} \,.
\end{equation}
In equilibrium, these two components are linked via the fluctuation-dissipation theorem,
\begin{equation}
\label{eq:fluct_diss}
 \Delta^{Ki} = \Delta^{Ri} (1-2f(\omega-\mu)) \,,
\end{equation}
where $f$ is the Fermi function and $\mu$ the chemical potential. %
For a \almostfull/\almostempty equilibrium bath the relation %
\begin{equation}
 \label{eq:RK_relation_FE}
 \Delta^{Ki}_{F/E} = \mp \Delta^{Ri}_{F/E} 
\end{equation}
follows from \eq{eq:fluct_diss} for $f \equiv 1$ (F) or $0$ (E), respectively.
We can, therefore, decompose
\begin{alignat*}{2}
 &\Delta^{Ki} ~&= \Delta^{Ki}_F + \Delta^{Ki}_E &= - \Delta^{Ri}_F + \Delta^{Ri}_E \,, \\
 &\Delta^{Ri} &&= \hphantom{-} \Delta^{Ri}_F + \Delta^{Ri}_E \,, 
\end{alignat*}
which gives
\begin{equation}
 \label{eq:FE_relation_RK}
 \Delta^{Ri}_{F/E} = \frac{\Delta^{Ri} \mp \Delta^{Ki}}{2} \,.
\end{equation}
Note that \eqs{eq:fluct_diss} and \eqref{eq:RK_relation_FE} are equilibrium properties. 
Therefore, these are valid for any component of each one of the two (uncoupled) baths, $E$ and $F$, and in particular for the Green's function matrix.  
Moreover, a matrix inversion preserves these relations.
However, for a matrix $\boldsymbol{A}^\beta$, $\beta \in \{R,K\}$, such as the Green's function or self energy matrix, one has to
replace the imaginary part \eqref{dri} with the anti-Hermitian part, i.e.: 
\begin{equation}
\label{eq:matrix_ri_ki}
\begin{split}
 \boldsymbol{A}^{Ri} &= \frac{1}{2i} (\boldsymbol{A}^R - {\boldsymbol{A}^R}^\dagger) \\
 \boldsymbol{A}^{Ki} &= \frac{1}{4i} (\boldsymbol{A}^K - {\boldsymbol{A}^K}^\dagger) 
\end{split} 
\end{equation}
Notice that the Keldysh component $\boldsymbol{A}^K$ is anti-Hermitian anyway.
In this case \eqref{eq:RK_relation_FE} becomes
\begin{equation}
\label{baki}
 \boldsymbol{A}^{Ki}_{F/E} = \mp \boldsymbol{A}^{Ri}_{F/E} \,,
\end{equation}

Applying \eq{eq:matrix_ri_ki} to the  Green's function matrix of one of the two uncoupled baths (cf.~Eqs.~(40) and (41) in \tcite{do.nu.14}), 
\begin{align*}
 \left( \underline{\boldsymbol{G}}^{-1} \right)^R &= \omega \boldsymbol{I} - \boldsymbol{E} + i\left(\boldsymbol{\Gamma}^{(1)}+\boldsymbol{\Gamma}^{(2)} \right) \,, \\
 \left( \underline{\boldsymbol{G}}^{-1} \right)^K &= -2i \left(\boldsymbol{\Gamma}^{(2)}-\boldsymbol{\Gamma}^{(1)} \right) \,,
\end{align*}
results in
\begin{equation}
\begin{split}
 \left( \underline{\boldsymbol{G}}^{-1} \right)^{Ri} &= \boldsymbol{\Gamma}^{(1)}+\boldsymbol{\Gamma}^{(2)} \,,\\
 \left( \underline{\boldsymbol{G}}^{-1} \right)^{Ki} &= \boldsymbol{\Gamma}^{(1)}-\boldsymbol{\Gamma}^{(2)} \,. 
\end{split}
\end{equation}
Inserting this result further into \eq{baki}
yields that a \almostfull bath has $\boldsymbol{\Gamma}^{(1)} = 0$ and an \almostempty one $\boldsymbol{\Gamma}^{(2)} = 0$, as expected,
\begin{equation}
 \left( \underline{\boldsymbol{G}}^{-1} \right)^{Ri}_{F} = \boldsymbol{\Gamma}^{(2)} \,, \quad
 \left( \underline{\boldsymbol{G}}^{-1} \right)^{Ri}_{E} = \boldsymbol{\Gamma}^{(1)} \,. 
\end{equation}

Notice that this splitting procedure does not change the properties of the impurity.
Furthermore, it can be carried out also for an equilibrium bath or for a situation in which the leads are partially full or partially empty.
A crucial point is that in MPS, it is always convenient to split the baths in this way, because the entanglement is less severe, see \tcite{do.ga.15}.

\subsection{From equilibrium to nonequilibrium}
\label{subsec:EQ_bath_to_NEQ_bath}
\begin{figure}[h!]
\subfigure[]{
  \label{subfig:app_RL_EQ_1}
  \begin{minipage}[b]{\pw\columnwidth} %
    \centering \includegraphics[width=0.75\textwidth]{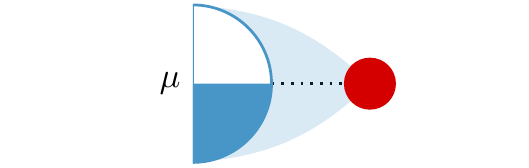}
  \end{minipage}
}
\subfigure[]{
  \label{subfig:app_FE_EQ_1}
  \begin{minipage}[b]{\pw\columnwidth} %
    \centering \includegraphics[width=0.75\textwidth]{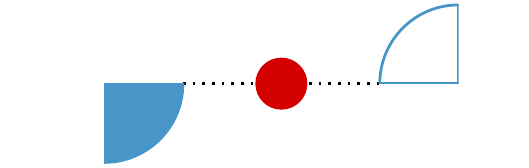}
  \end{minipage}
}
\caption{\subref{subfig:app_RL_EQ_1} Impurity (red sphere) coupled to a partially filled bath (semicircle) at chemical potential $\mu$. \subref{subfig:app_FE_EQ_1} The same 
hybridization function can be obtained by coupling the impurity to a \almostfull and \almostempty bath with appropriate DOS.
}
\label{fig:bath_EQ_1}
\end{figure}

As discussed, we start by fitting a bath in equilibrium and then we split it into a \almostfull and an \almostempty one, see \fig{fig:bath_EQ_1}. 
In fact, it turns out that such a geometry  naturally comes out for a chain geometry fit.

For the situation depicted in \fig{subfig:app_FE_EQ_1} the fit produces the following Lindblad matrices, assuming PH symmetry,
\begin{equation}
\label{eq:orig_Emat}
 \boldsymbol{E} = 
 \left(
  \begin{array}{c|c|c}
   \tilde{\boldsymbol{E}}^{\tau}  & \begin{array}{c} 0 \\ t \end{array} & 0 \\ \hline  
   \begin{array}{cc} 0 & t \end{array} & \varepsilon_f & \begin{array}{cc} t & 0 \end{array} \\ \hline 
   0 & \begin{array}{c} t \\ 0 \end{array} & \tilde{\boldsymbol{E}}
 \end{array}
 \right)
\end{equation}
and 
\begin{equation*}
 \boldsymbol{\Gamma}^{(1)} =
 \left(
  \begin{array}{c|c|c}
   \tilde{\boldsymbol{\Gamma}}^{\tau}  & 0 & 0 \\ \hline 
   0 & 0 & 0 \\ \hline 
   0 & 0 & 0
 \end{array}
 \right) \,,
 \quad
 \boldsymbol{\Gamma}^{(2)} =
 \left(
  \begin{array}{c|c|c}
   0 & 0 & 0 \\ \hline  
   0 & 0 & 0 \\ \hline 
   0 & 0 & \tilde{\boldsymbol{\Gamma}}
 \end{array}
 \right) \,.
\end{equation*}
Here, $\tilde{\boldsymbol{E}}$ and $\tilde{\boldsymbol{\Gamma}}$ are $N_B/2 \times N_B/2$ block matrices and each matrix $\boldsymbol{A}^{\tau}$ is $\boldsymbol{A}$ with the order of indices inverted and different signs, see Eq.~(27) in \tcite{do.so.17}, for the exact relations.
For MPS, $\tilde{\boldsymbol{E}}$ and $\tilde{\boldsymbol{\Gamma}}$ should be tridiagonal, which corresponds to having nearest-neighbor hoppings and $\Gamma$ terms only.
The retarded hybridization function of, for instance,  the \almostfull bath is then given by
\begin{equation}
 \label{eq:hyb_R_F}
 \Delta^R_F(\omega) = t^2 \bar{\gamma}^R(\omega) 
\end{equation}
with the boundary Green’s function
\begin{equation}
 \label{eq:gamma_R}
 \bar{\gamma}^R(\omega) = \left[\left(\omega \boldsymbol{I} - \tilde{\boldsymbol{E}} + i \tilde{\boldsymbol{\Gamma}} \right)^{-1} \right]_{11}
\end{equation}
and the Keldysh hybridization function $\Delta^K_F(\omega)$ is fixed by \eq{eq:RK_relation_FE}.
The result for the \almostempty bath follows from PH symmetry.

\begin{figure}[h!]
\subfigure[]{
  \label{subfig:app_RL_NEQ}
  \begin{minipage}[b]{\pw\columnwidth} %
    \centering \includegraphics[width=0.75\textwidth]{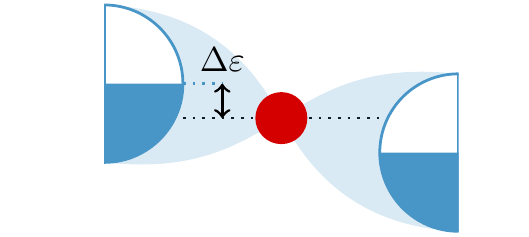}
  \end{minipage}
}
\subfigure[]{
  \label{subfig:app_FE_NEQ}
  \begin{minipage}[b]{\pw\columnwidth} %
    \centering \includegraphics[width=0.75\textwidth]{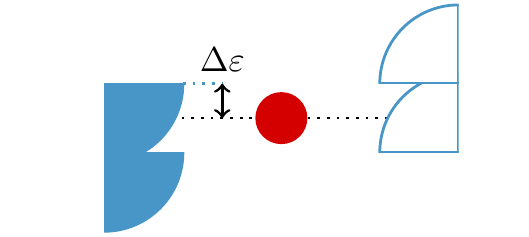}
  \end{minipage}
}
\caption{\subref{subfig:app_RL_NEQ} Impurity (red sphere) coupled to a partially filled left bath and a partially filled right bath (semicircles), whose chemical potentials differ by $2 \Delta \varepsilon$. \subref{subfig:app_FE_NEQ} The same situation with two \almostfull (blue) and two \almostempty (white) baths.
}
\label{fig:bath_NEQ}
\end{figure}

Instead of the equilibrium situation in \fig{subfig:app_RL_EQ_1}, we would now like to represent a nonequilibrium one, as depicted in \fig{subfig:app_RL_NEQ}.
If the total DOS is fixed, this is obtained by reducing the hoppings to the impurity by $1/\sqrt{2}$ and by doubling the number of bath sites and shifting their on-site energies by $\pm \Delta\varepsilon$.
Then \fig{subfig:app_FE_EQ_1} schematically becomes \fig{subfig:app_FE_NEQ}, which can no longer be represented in a chain geometry (with tridiagonal matrices).
The matrix in \eq{eq:orig_Emat} becomes
\begin{equation}
\label{eq:mod_Emat}
 \boldsymbol{E}^\prime = 
 \left(
  \begin{array}{c|c|c|c|c}
   \tilde{\boldsymbol{E}}^{\tau} + \Delta\varepsilon \boldsymbol{I}  & 0 & \begin{array}{c} 0 \\ t^\prime \end{array} & 0 & 0\\ \hline  
   0 & \tilde{\boldsymbol{E}}^{\tau} - \Delta\varepsilon \boldsymbol{I}  & \begin{array}{c} 0 \\ t^\prime \end{array} & 0 & 0\\ \hline 
   \begin{array}{cc} 0 & t^\prime \end{array} & \begin{array}{cc} 0 & t^\prime \end{array} & \varepsilon_f & \begin{array}{cc} t^\prime & 0 \end{array} & \begin{array}{cc} t^\prime & 0 \end{array} \\ \hline
   0 & 0 & \begin{array}{c} t^\prime \\ 0 \end{array} & \tilde{\boldsymbol{E}} - \Delta\varepsilon \boldsymbol{I} & 0 \\ \hline
   0 & 0 & \begin{array}{c} t^\prime \\ 0 \end{array} & 0 & \tilde{\boldsymbol{E}} + \Delta\varepsilon \boldsymbol{I}
 \end{array}
 \right)
\end{equation}
\\
with $t^\prime = t/\sqrt{2}$ and, correspondingly, $\boldsymbol{\Gamma}^{(1)}$ and $\boldsymbol{\Gamma}^{(2)}$.
In this situation, \eq{eq:hyb_R_F} still holds, but instead of \eq{eq:gamma_R}, we have 
\begin{equation*}
 \Delta^R_F(\omega) = \frac{t^2}{2} \left( \bar{\gamma}^R(\omega + \Delta\varepsilon) + \bar{\gamma}^R(\omega - \Delta\varepsilon) \right) \,.
\end{equation*}
However, the matrix \eqref{eq:mod_Emat} is not suitable for MPS, as it is not tridiagonal.
To make progress, we observe that $\Delta^R_F(\omega)$ can be obtained by considering the following matrix in block form
\begin{equation*}
\label{eq:mod_Emat_part}
 \boldsymbol{h}^\prime = 
 \left(
  \begin{array}{c|c|c}
   0 & \begin{array}{cc} t^\prime & 0 \end{array} & \begin{array}{cc} t^\prime & 0 \end{array} \\ \hline
   \begin{array}{c} t^\prime \\ 0 \end{array} & \tilde{\boldsymbol{E}} - i\tilde{\boldsymbol{\Gamma}} - \Delta\varepsilon \boldsymbol{I} & 0 \\ \hline
   \begin{array}{c} t^\prime \\ 0 \end{array} & 0 & \tilde{\boldsymbol{E}} - i\tilde{\boldsymbol{\Gamma}} + \Delta\varepsilon \boldsymbol{I}
 \end{array}
 \right)
\end{equation*}
and employing Dyson's equation, %
\begin{equation}
 \Delta^R_F(\omega) = \omega - \frac{1}{[(\omega \boldsymbol{I} - \boldsymbol{h}^\prime)^{-1}]_{11}} \,.
\end{equation}
For MPS we need a tridiagonal form, as discussed above.
This can be achieved with a Bi-Lanczos transformation.
All we need is that $[(\omega \boldsymbol{I} - \boldsymbol{h}^\prime)^{-1}]_{11}$ remains invariant.
The transformation is produced by a matrix (here the upper block is $1\times 1$ and the lower is $N_B\times N_B$)
\begin{equation}
 \boldsymbol{U} =
 \left(
  \begin{array}{c|c}
   1  & 0 \\ \hline  
   0 & \tilde{\boldsymbol{U}} 
 \end{array}
 \right) \,,
\end{equation}
where $U$ is, in general, non-unitary, yielding
\begin{equation*}
\begin{split}
 \boldsymbol{h}^{\prime\prime} &= \boldsymbol{U}^{-1} \boldsymbol{h}^\prime \boldsymbol{U} \\
 &= 
 \left(
  \begin{array}{c|c}
   0 & \begin{array}{cc} t^{\prime\prime} & 0 \end{array} \\ \hline
   \begin{array}{c} t^{\prime\prime} \\ 0 \end{array} & \boldsymbol{H}^{\prime\prime}
 \end{array}
 \right) \,.
\end{split}
\end{equation*}
Here, the non-Hermitian tridiagonal matrix $\boldsymbol{H}^{\prime\prime}$ identifies the new parameters of the \almostfull (F) bath, 
\begin{equation}
 \begin{split}
  \tilde{\boldsymbol{E}}^{\prime\prime} &\equiv \frac{{\boldsymbol{H}^{\prime\prime}}^\dagger   + \boldsymbol{H}^{\prime\prime}}{2} \,,\\
  \tilde{\boldsymbol{\Gamma}}^{\prime\prime} &\equiv \frac{{\boldsymbol{H}^{\prime\prime}}^\dagger - \boldsymbol{H}^{\prime\prime}}{2i} \,,
 \end{split}
\end{equation}
while the ones of the \almostempty (E) bath are obtained by PH symmetry, see Eq.~(27) in \tcite{do.so.17}.

Note that, since $\tilde{\boldsymbol{U}}$ is not unitary, $\tilde{\boldsymbol{E}}^{\prime\prime}$ and $\tilde{\boldsymbol{\Gamma}}^{\prime\prime}$ are not simply obtained by transforming $\tilde{\boldsymbol{E}}$ and $\tilde{\boldsymbol{\Gamma}}$, separately.
This can, and in our case does, produce $\tilde{\boldsymbol{\Gamma}}^{\prime\prime}$ that are not semi-positive definite, as should be required for the Lindblad equation.
Still, the steady state we obtain is stable and the spectral functions turn out to be causal.
The reason is that the new parameters originate from semi-positive definite matrices.
\section{Symmetry considerations and error estimation}
\label{sec:error_estimation}
In principle, we can calculate four Green's functions individually, $G^\alpha_\sigma$ with $\sigma \in \{\uparrow,\downarrow\}$ and $\alpha \in \{<,>\}$.
The system, though, is PH symmetric, which relates the lesser and the greater Green's function to each other, and it is spin symmetric.
Therefore, the following relations must be fulfilled,
\begin{align}
 G^<_\sigma(x) &= -G^>_\sigma(-x)\,, \label{eq:ph_symm}\\
 G^\alpha_\uparrow(x) &= G^\alpha_\downarrow(x) \,, \label{eq:spin_symm}
\end{align}
for $x$ being either $t$ or $\omega$.
This reduces the number of actually independent Green’s functions to only one.
Thus, in order to obtain the spectral function, for example, it is in principle sufficient to calculate only one $G^\alpha_\sigma$, then construct $G^{\bar{\alpha}}_\sigma$ with $\bar{\alpha} \neq \alpha$ from \eq{eq:ph_symm} and evaluate \eq{eq:A_from_G}.
We refer to this as \textbf{protocol 1}.

\begin{figure}[h]
  \begin{minipage}[b]{\pw\columnwidth} %
    \centering \includegraphics[width=1\textwidth]{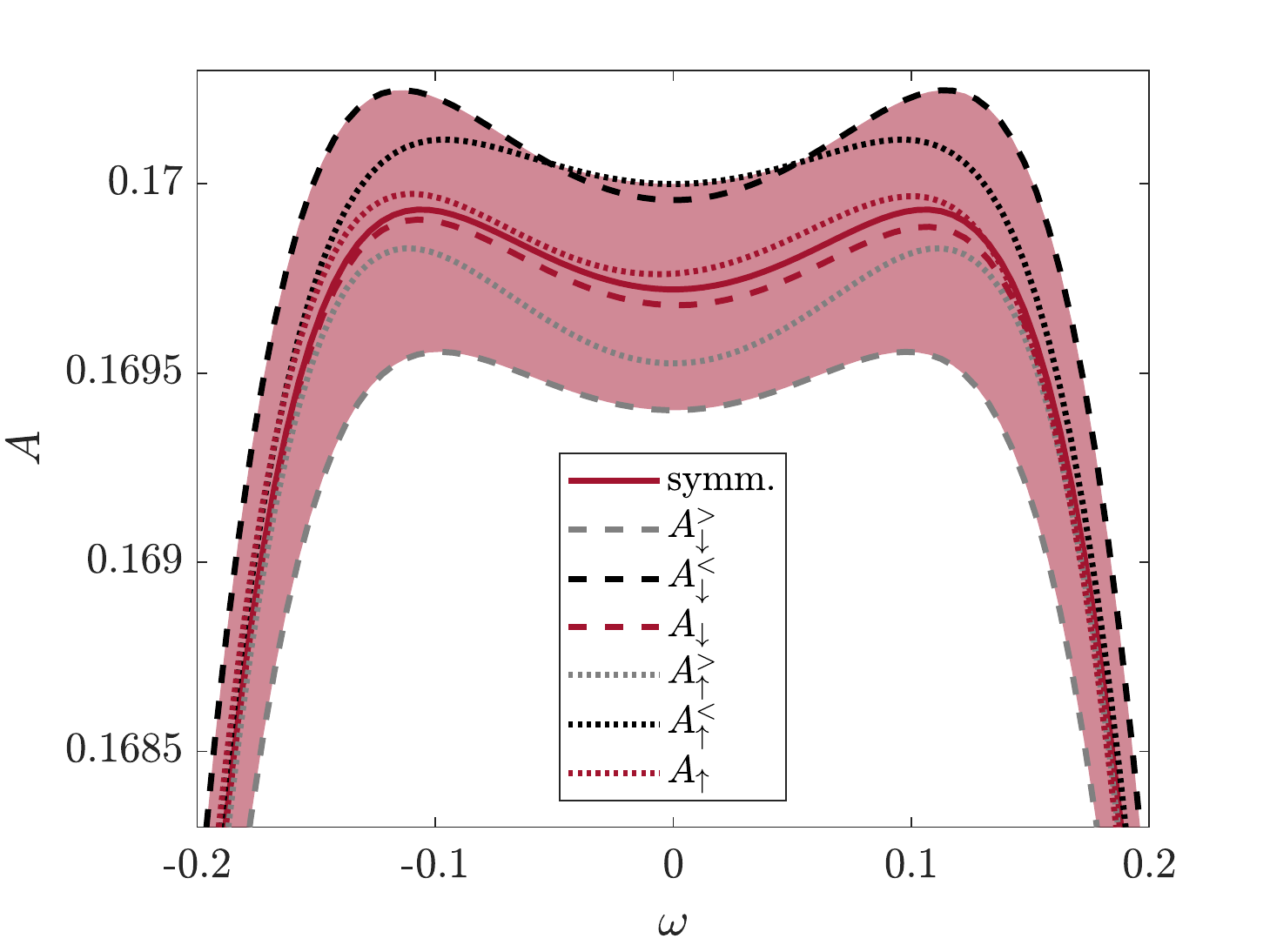} 
  \end{minipage}
\caption{
Auxiliary spectral functions $A(\omega)$ obtained from different raw data for symmetry considerations and error estimation, see text.
}
\label{fig:neq_symm_1}
\end{figure}

However, if we calculate $G^\alpha_\sigma$ with AMEA employing MPS, the symmetry relations, \eqs{eq:ph_symm}-\eqref{eq:spin_symm}, are not exactly fulfilled.
This is, due to the approximations within the MPS calculation, more specifically, due to the truncation and Suzuki-Trotter errors.
\fig{fig:neq_symm_1} shows the consequences of these violations at the example of the spectral function.

We can see that the spectral functions determined from only one $G^\alpha_\sigma$, according to protocol 1, are symmetric by construction, $A^\alpha_\sigma(\omega) = A^\alpha_\sigma(-\omega)$, but they differ from each other, $A^\alpha_\sigma(\omega) \neq A^{\bar{\alpha}}_{\bar{\sigma}}(\omega)$ for $\alpha \neq \bar{\alpha}$ and $\sigma \neq \bar{\sigma}$.
The area enclosed by the four different solutions is color-shaded and the solid curve in the center is the average of these solutions, which we call symmetrized spectral function in this paper.
The deviations of the borders of the shaded area from the symmetrized spectral function can be used as a measure for the symmetry errors throughout the MPS calculation.

In this figure, we can also see the spectral functions naively determined from two Green's functions, $G^<_\sigma$ and $G^>_\sigma$, by evaluating \eq{eq:A_from_G} directly, without enforcing PH symmetry. 
We refer to this as \textbf{protocol 2}.
These spectral functions are not exactly symmetric, $A_\sigma(\omega) \neq A_\sigma(-\omega)$, as discussed above, but they are close to the symmetrized spectral function and they lie almost entirely within the shaded area for almost all bias voltages (except $\phi = 0.8$ and $\phi = 1$).

Throughout this paper, we display also other, in principle symmetric, quantities as symmetrized curves with errors in the form of color-shaded areas, obtained by protocol 1.
Specifically, the self energy and the differential conductance are represented in this way, see \figs{fig:eq_expos}, \ref{fig:neq_Kondo} and \ref{fig:neq_LM}. 
For the differential conductance, we also consider deviations arising by protocol 2 and plot the corresponding errors separately, as bars, in addition to the shaded area, see \figs{subfig:G_neq_Kondo} and \ref{subfig:G_neq_LM}. 
For the other quantities these errors lie almost entirely within the shaded area, anyway, and their inclusion does not make any difference.
The differential conductance, though, being obtained as a numerical derivative of these quantities by \eqs{eq:curr_formula} and \eqref{eq:G_general_formula}, is more sensitive to deviations.
\newpage
\bibliographystyle{/afs/itp.tugraz.at/user/arrigoni/bibtex/prsty} 
\bibliography{/afs/itp.tugraz.at/user/arrigoni/bibtex/references_database,bibs/refs_Delia}

\end{document}